\documentclass[a4paper, onecolumn, pra, superscriptaddress, notitlepage, accepted=2023-01-12]{quantumarticle}
\pdfoutput=1
\usepackage{setspace}

\usepackage{amsmath}
\usepackage[utf8]{inputenc}
\usepackage{graphicx}
\usepackage{tikz}
\usepackage{physics}
\usepackage{amsmath}
\usepackage{epstopdf}
\usepackage{amssymb}
\usepackage{mathtools}
\usepackage{multirow}
\usepackage{amsthm}
\usepackage[toc,page]{appendix}
\usepackage{natbib}
\usepackage{graphicx}

\newcommand{\proj}[1]{|#1\rangle\!\langle #1|}

\usepackage[colorlinks]{hyperref}
\AtBeginDocument{%
  \hypersetup{
    citecolor=blue,
    linkcolor=blue,
    urlcolor=blue}}

\usepackage[a4paper, total={6.2in, 9in}]{geometry}

\usepackage{amssymb}
\usepackage[shortlabels]{enumitem}
\usepackage{qcircuit}
\usepackage[english]{babel}
\usepackage{algorithm}
\usepackage{dsfont}
\usepackage{qcircuit}
\usepackage{color} 
\usepackage[framemethod=TikZ]{mdframed}
\usepackage{algorithm}
\usepackage{algpseudocode}
\usepackage{braket}
\usepackage{etoolbox}
\apptocmd{\sloppy}{\hbadness 10000\relax}{}{}
\usepackage[labeled]{multibib}
\newcites{S}{Supplementary Material References}
\usepackage{url}

\usepackage{tikz}
\usepackage{graphicx}
\usetikzlibrary{positioning}
\usepackage{fancyhdr}
\usepackage[utf8]{inputenc}

\usepackage{pgfplots}
\pgfplotsset{compat=newest}
\usepgfplotslibrary{groupplots}
\usepgfplotslibrary{dateplot}

\usepackage{graphicx}
\usepackage{subcaption}

\usepackage{tocloft}
\usepackage{lastpage}

\begin{document}

\title{Variational quantum algorithm for unconstrained black box binary optimization: Application to feature selection}

\author{Christa Zoufal}
\affiliation{IBM Quantum, IBM Research Europe -- Zurich}
\author{Ryan V. Mishmash}
\affiliation{IBM Quantum, Almaden Research Center -- Almaden}
\author{Nitin Sharma}
\affiliation{PayPal -- San Jose}
\author{Niraj Kumar}
\affiliation{PayPal -- San Jose}
\thanks{While working on this project, N.K. was affiliated with PayPal. He is currently affiliated with JP Morgan Chase $\&$ Co.}
\author{Aashish Sheshadri}
\affiliation{PayPal -- San Jose}
\author{Amol Deshmukh}
\affiliation{IBM Quantum, Thomas J. Watson Research Center -- Yorktown Heights}
\author{Noelle Ibrahim}
\affiliation{IBM Quantum, Thomas J. Watson Research Center -- Yorktown Heights}
\author{Julien Gacon}
\affiliation{IBM Quantum, IBM Research -- Zurich}
\affiliation{Institute of Physics, \'Ecole Polytechnique F\'ed\'erale de Lausanne (EPFL)}
\author{Stefan Woerner}
\affiliation{IBM Quantum, IBM Research -- Zurich}

\begin{abstract}

We introduce a variational quantum algorithm to solve unconstrained black box binary optimization problems, i.e., problems in which the objective function is given as black box.
This is in contrast to the typical setting of quantum algorithms for optimization where a classical objective function is provided as a given Quadratic Unconstrained Binary Optimization problem and mapped to a sum of Pauli operators.
Furthermore, we provide theoretical justification for our method based on convergence guarantees of quantum imaginary time evolution.

To investigate the performance of our algorithm and its potential advantages, we tackle a challenging real-world optimization problem: \emph{feature selection}. This refers to the problem of selecting a subset of relevant
features to use for constructing a predictive model such as fraud detection. Optimal feature selection---when formulated in terms of a generic loss function---offers little structure on which to build classical heuristics, thus resulting primarily in ‘greedy methods’. This leaves room for (near-term) quantum algorithms to be competitive to classical state-of-the-art approaches.
 We apply our quantum-optimization-based feature selection algorithm, termed VarQFS, to build a predictive model for a credit risk data set with $20$ and $59$ input features (qubits) and train the model using quantum hardware and tensor-network-based numerical simulations, respectively. We show that the quantum method produces competitive and in certain aspects even better performance compared to traditional feature selection techniques used in today's industry.
\end{abstract}

\maketitle

\section{Introduction}
In recent years, quantum computing has emerged as a promising platform on which to apply heuristic approaches to solve computationally expensive combinatorial optimization tasks \cite{Farhi2014, Moll2017, Barkoutsos2019, Egger2021, Gacon_2020QSBO, AmaroEPJVariationalQAJobShopScheduling22, Amaro_2022FilteringVariationalQuantumAlgorithms, DacremaFSQuantumAnnealers22}.
Most of these approaches have in common that they consider a Quadratic Unconstrained Binary Optimization (QUBO) problem. 
In this work, we introduce a variational quantum optimization algorithm that can be applied to the much broader class of unconstrained black box binary optimization problems.
More explicitly, the algorithm takes a black box optimization problem $f\left(\boldsymbol{x}\right)$ as input and returns the optimal solution $\boldsymbol{x}_{\text{opt}} = \min_{\boldsymbol{x}\in\set{0,1}^n} f\left(\boldsymbol{x}\right)$ or a good approximation thereof. The solution space of these optimization problems is typically exponential in $n$.
Mapping the solution space to the computational basis states of a parameterized quantum state with $n$ qubits, allows an algorithm definition where the search for $\boldsymbol{x}_{\text{opt}}$ is conducted by optimization of the state parameters.

It should be noted that QUBOs are a special case of the considered problem class, and that one might be able to find a QUBO that represents or at least approximates a black box binary optimization problem. However, in general this is not possible efficiently.
Furthermore, the respective objective function implicitly defines a diagonal Hamiltonian, whose ground state corresponds to the optimal solution of the considered optimization problem.
Hence, finding the solution of the given black box optimization may be approached with various ground state approximation techniques such as the variational quantum eigensolver (VQE) \cite{peruzzo2014variational, Amaro_2022FilteringVariationalQuantumAlgorithms}.
In this work, we search for the problem solution with a method that, firstly, (in contrast to standard VQE) comes with a strong theoretical justification for finding a solution close to the ground state and, secondly, has promising scaling behavior.
More specifically, we employ an efficient stochastic approximation to the quantum natural gradients algorithm which is closely related to quantum imaginary time evolution \cite{VarSITEMcArdle19}.
The key component of quantum imaginary time evolution is the continuous suppression of the sampling probabilities of all excited states while amplifying the sampling probability of the ground state. Therefore, the only state that remains after sufficient evolution time is the ground state (and similarly for degenerate ground states). 
A comparable concept for optimization based on damping the amplitude magnitudes of large energy eigenstates is suggested in a VQE setting in \cite{Amaro_2022FilteringVariationalQuantumAlgorithms}. The authors employ a set of filtering operators which applied to the quantum state increase the magnitude of the amplitudes of lower eigenstates. At each optimization step, the algorithm aims to find a set of parameters with which the variational state approximates the filter operator mapped state as good as possible. For an exponential filter operator, this approach can be interpreted as a quantum imaginary time evolution approximation. While the authors of \cite{Amaro_2022FilteringVariationalQuantumAlgorithms} focus on QUBO based problems, our approach now shows how to employ the idea of damping the amplitude magnitues of large energy eigenstates for a generalized optimization setting.

To illustrate the potential of this algorithm in a specific context, we are going to focus our analysis and discussion on the example of \emph{feature selection}, a challenging and practically relevant optimization problem.
Feature selection generically refers to the process of selecting a subset of relevant features and/or eliminating redundant or irrelevant features from a given data set. To that end, one aims at keeping only a subset of features that is most relevant for a given learning task while discarding the rest. As the cardinality of the feature space grows, the problem of determining the best feature subset
depicts exponential growth.

We devise and test the variational quantum optimization algorithm for feature selection---termed \emph{variational quantum feature selection} (VarQFS)---in which a parameterized quantum circuit is trained to provide \textit{good} feature subsets.
Hereby, a subset is referred to as \textit{good} if the respective features lead to a model which performs well in terms of a predefined metric. The suitability or fitness of a subset of features depends not only
on the underlying data distribution, but also on the type of algorithm and the associated hyperparameters
and can be measured using different criteria depending on the response type.
Furthermore, we investigate the application and resulting performance of VarQFS on a publicly available real-world credit risk data set \cite{Dua:2019_MLRepo}. Our VarQFS algorithm employs a promising, scalable optimizer, i.e., Quantum Natural SPSA (QN-SPSA) \cite{gacon2021simultaneous} to train a parameterized quantum circuit. The approach facilitates performance investigation as well as feasibility demonstrations on systems up to $59$ qubits with a matrix product state (MPS) simulator and on systems with $20$ qubits on actual quantum hardware. The results reveal that the suggested approach is not only able to compete with state-of-the-art classical methods but it even manages to outperform them in some of the presented examples. In fact, VarQFS is able to find feature subsets that lead to better fitness results with respect to a predefined metric compared to the classical benchmarks. 
 
The rest of the paper is structured as follows. First, the classical feature selection methods are discussed in Sec.~\ref{sec:trad_fs}.
Next, we introduce the variational quantum optimization algorithm which helps to find heuristic solutions to unconstrained black box binary optimization problems and discuss its use for VarQFS in Sec.~\ref{sec:VarQFS}. Then, we present the simulation and hardware experiment settings and results in Sec.~\ref{sec:results}. Finally, a conclusion and an outlook are given in Sec.~\ref{sec:conclusion}.

\section{Feature Selection}
\label{sec:trad_fs}

Feature selection helps to simplify learning problems respectively to improve training properties.
Given a training data set with thousands of features, it may become expensive or even infeasible to train a machine learning (ML) model on this data.
Feature selection describes the task of selecting a subset of features which are then used to train a given model. The main goal of this approach is speeding up the model training process, reducing memory costs during the training, and avoiding overfitting.
Furthermore, it is not always the case that every feature is important for the training. Therefore, feature selection can also help to increase the interpretability of the model, and to improve generalizability.
However, finding the optimal subset of a set of features for a given data set is a combinatorial optimization problem that is NP-hard \cite{BLUM1997245FeatureSelection}.
Typical classical feature selection heuristics investigate the features independently or consider at most pair-wise correlations.
Generally, feature selection represents an optimization problem which offers little structure to be exploited by classical techniques and
thus offers an interesting test-bed to study potential advantages of quantum algorithms.

The main categories of classical feature selection techniques are given as follows: \textit{model-based} methods, \textit{filter} methods, and \textit{wrapper} methods \cite{kuhn2013applied}. 
\textit{Model-based} techniques provide methods to rank features based on the feature importance score generated as part of the learning process. The \textit{filter} methods use statistical techniques to evaluate the relationship between each feature and the output variable. 
Such methods typically assess feature fitness with statistical tests or thresholds that filter out features which are unable to meet the thresholds.
Finally, \textit{wrapper} methods assess the performances of many models composed of different features and select those features that perform best according to a given performance metric. These methods are usually iterative or sequential in nature and involve evaluating performances using multiple sets of features, and are thus computationally expensive; nevertheless, they do perform better than the other two classes of methods in practise. One of the most familiar examples of wrapper methods is a technique called \textit{Recursive Feature Elimination} (RFE) \cite{guyon2002gene}. 

RFE employs an iterative backward elimination process. It evaluates the performance of an underlying model on successively smaller subsets of features. The training process of the model begins by including all the raw features. The next step is the elimination of the least important features as determined by the feature ranking scores provided by the underlying model, such as the 
shift in model scores upon feature nullification,
weights or feature importances. These two steps are recursively implemented until the predefined number of features is eventually reached or the best performance on validation data is achieved.

An extension of RFE involves an automatic search over the optimal number of features to be selected using a cross-validation procedure and is referred to as RFECV. 
Both RFE and RFECV involve a greedy, iterative procedure for selecting a subset of features which is prone to not properly accounting for the correlations between multiple features.
Furthermore, the feature importance ranking, which is at the core of RFE and RFECV, may not be unique when there exists multicollinearity among features. Several folds could be used to overcome this problem. Each fold may produce a slightly different performance ranking of features with some collective decision being made to produce stability. Nevertheless, the number of folds is a hyperparameter of the algorithm and there is no guarantee that a different selection of the folds would not produce a slightly different answer. 

We would like to point out that one can also formulate a feature selection problem in terms of a QUBO, which could be solved, e.g., with VQE or QAOA, see \cite{MueckeQFS22}, or adiabatic quantum computing, see \cite{1Qubit_QuantumFeatureSelection17}. However, the QUBO based optimization leads to an intrinsically different problem setting than considered here.
More specifically, the construction of the respective QUBO is based on correlation coefficients between each individual feature of the various data points and the corresponding labels as well as on correlation coefficients between every pair of features of the data.
However, the evaluation of those correlation coefficients requires a significant overhead. Furthermore, this method only enables to capture up to pair-wise correlations.

\section{Variational Quantum Optimization}
\label{sec:VarQFS}
This section describes a variational quantum optimization algorithm which can be used to tackle a large class of optimization problems with feature selection being just one such example.
To that end, we prepare a quantum state with a parameterized quantum circuit, identify its basis states with states of the solution space, and define the loss function according to the optimization task at hand. The model is then trained with the goal of maximizing the probability to sample the optimal solution from the parameterized quantum state.
We present the details of our scheme in generality in Sec.~\ref{sec:optimization}; its application to VarQFS is described in Sec.~\ref{sec:VarQFS_application}.

\subsection{Optimization}
\label{sec:optimization}

Suppose an unconstrained binary optimization problem 
\[
\min_{\boldsymbol{x} \in \{0, 1\}^n} f(\boldsymbol{x}),
\]
where the objective function $f: \{0, 1\}^n \rightarrow \mathbb{R}$ with $|f(\boldsymbol{x})| = \mathcal{O}\left(poly(n)\right)$, may be given as a black box.
Then, we use an $n$-qubit parameterized quantum circuit $U\left(\boldsymbol{\theta}\right)$ with $\boldsymbol{\theta}=\set{\theta_0, \ldots, \theta_{k-1}}$
to prepare a parameterized quantum state:
\begin{eqnarray}
\ket{\psi\left(\boldsymbol{\theta}\right)} = U\left(\boldsymbol{\theta}\right)\ket{0}^{\otimes n} = \sum\limits_{j=0}^{2^n-1}e^{-i\phi_j(\boldsymbol{\theta})}\sqrt{p_j\left(\boldsymbol{\theta}\right)}\ket{j},
\end{eqnarray}
where $e^{-i\phi_j}$ corresponds to an arbitrary phase and $p_j\left(\boldsymbol{\theta}\right)$ is the probability to sample $\ket{j}$.
Further, the basis states $\ket{j}$ are identified with possible solutions $\boldsymbol{x}_j$ of the optimization problem.
While the sampling probabilities $p_j\left(\boldsymbol{\theta} \right)$ and basis states $\ket{j}$ are important for this context, the phase terms $e^{-i\phi_j}$ can be neglected. Hence, it suffices to choose an ansatz that acts in the real space of the amplitudes.

Now, the goal of the variational quantum optimization is to maximize the probability to sample solutions $\boldsymbol{x}_j$ corresponding to good objective values $f(\boldsymbol{x}_j)$ from $\ket{\psi\left(\boldsymbol{\theta}\right)}$ by optimizing a loss function
\begin{align}
\label{eq:loss}
    L\left(\boldsymbol{\theta}\right) = \sum_j p_j\left(\boldsymbol{\theta}\right) f\left(\boldsymbol{x}_j\right),
\end{align}
with respect to $\boldsymbol{\theta}$.

An important aspect of the variational quantum optimization presented above is that it is not necessary to have an explicit Pauli-operator representation of the objective function. Instead, to evaluate Eq.~\eqref{eq:loss} we must generate samples (shots) according to the sampling probabilities $p_j(\boldsymbol{\theta})$, calculate the values $f(\boldsymbol{x}_j)$ via the given black box function,
and finally sum up the results.
Although the loss function consists of a sum over an exponential number of terms, it can be approximated well with a polynomial number of measurements. Consider that $N_{\text{shots}}$ number of samples $\boldsymbol{x}_s\left(\boldsymbol{\theta}\right)$ are independently
drawn from the quantum state $\ket{\psi\left(\boldsymbol{\theta}\right)}$ and used to evaluate an empirical estimate 
\begin{align}
    \tilde{L}\left(\boldsymbol{\theta}\right) = \frac{1}{N_{\text{shots}}}\sum\limits_{s=0}^{N_{\text{shots}}-1} f\left(\boldsymbol{x}_s\left(\boldsymbol{\theta}\right)\right)
\end{align}
to the loss function. 
Then, Hoeffding's inequality \cite{Hoeffding63} implies that
\begin{align}
    P\left(\lvert L\left(\boldsymbol{\theta}\right) - \tilde{L}\left(\boldsymbol{\theta}\right)\rvert \geq \epsilon \right) &\leq 2\exp\left({\frac{-2\epsilon^2N_{\text{shots}}}{\left(f^{\text{max}}-f^{\text{min}}\right)^2}}\right),
\end{align}
for $\epsilon>0$ and $f^{\text{min}} \leq f\left(\boldsymbol{x}_j\right) \leq f^{\text{max}},\quad\forall \boldsymbol{x}_j \in \set{0,1}^n$. By setting $P\left(\lvert L\left(\boldsymbol{\theta}\right) - \tilde{L}\left(\boldsymbol{\theta}\right)\rvert \geq \epsilon \right) =: \gamma$, we can directly find that
\begin{align}
    \epsilon \leq \alpha\left(\gamma\right) \sqrt{\frac{1}{N_{\text{shots}}}} \:,
\end{align}
It should further be noted that this optimization task can be written as a ground state problem with respect to an implicitly defined diagonal Hamiltonian
\begin{align}
    \min_{\boldsymbol{\theta}} \bra{\psi\left(\boldsymbol{\theta}\right)}H\ket{\psi\left(\boldsymbol{\theta}\right)},
\end{align}
for 
\begin{align}
\label{eq:diag_h}
  H=  \begin{pmatrix}
    f(\boldsymbol{x}_0) & & \\
    & \ddots & \\
    & & f(\boldsymbol{x}_{2^n-1})\\
    \end{pmatrix}.
\end{align}

Loss functions of this type are typically highly non-convex \cite{Huembeli_2021Hessian_Loss}. It is therefore of great importance to choose the right optimizer.
We focus on the application of an optimizer which is promising to achieve high performance for quantum problems \cite{yamamoto2019natural, lopatnikova2021quantum, QuantumStatLearningQWNGBecker21} using information-geometry-aware parameter training: \textit{quantum natural gradient} (QNG) \cite{Stokes_2020QNG}.
Standard-gradient-based optimizers update the parameters by searching for the steepest descent in the Euclidean geometry of the loss landscape.
Natural-gradient-based optimizers \cite{amari_why_1998}, on the other hand, focus on the natural geometry of the parameters.
 
Given that the optimization task corresponds to a ground state problem, exact QNG-based optimization is equivalent to variationally approximating quantum imaginary time evolution \cite{VarSITEMcArdle19}. Exact quantum imaginary time evolution is guaranteed to converge to the ground state for run time $t\rightarrow \infty$ if the initial state has a non-zero overlap.
With an adequately expressive ansatz, sufficiently small (adaptive) step size, and enough shots, a QNG-based optimizer follows the evolution of imaginary time in its convergence path to the global optimum. Although these conditions are difficult to suffice in practice, this gives theoretical motivation for the approach presented here. We would also like to point out that the approximation error induced by VarQITE can be efficiently bounded as is discussed in \cite{zoufal2021error}. These bounds may, then, be used to iteratively improve the variational implementation.

Formally, a QNG represents a gradient that is rescaled by the quantum geometric tensor---which in turn is proportional to the quantum Fisher information matrix (QFI) \cite{QFIMBraunstein94, meyer2021fisher}.
To compute a QNG, we first need to evaluate the gradient of the loss function $\nabla_{\boldsymbol{\theta}} L\left(\boldsymbol{\theta}\right)$.
Next, the QFI, ${\mathcal{F}^Q}$, is evaluated with its entries being given by
\begin{eqnarray}
\label{eq:qfi}
\mathcal{F}_{pq}^Q = 4\text{Re}\left(\frac{\partial \bra{\psi(\boldsymbol{\theta})}}{\partial \theta_p}\frac{\partial \ket{\psi(\boldsymbol{\theta})}}{\partial \theta_q} - \frac{\partial \bra{\psi(\boldsymbol{\theta})}}{\partial \theta_p}\proj{\psi(\boldsymbol{\theta})}\frac{\partial \ket{\psi(\boldsymbol{\theta})}}{\partial \theta_q}\right). 
\end{eqnarray}
The QNG finally corresponds to
\begin{align}
    {\Big(\frac{1}{4}\mathcal{F}^Q\Big)}^{-1}\nabla_{\boldsymbol{\theta}} L\left(\boldsymbol{\theta}\right) = {\Big(\frac{1}{4}\mathcal{F}^Q\Big)}^{-1}\sum_j \nabla_{\boldsymbol{\theta}} p_j\left(\boldsymbol{\theta}\right) f\left(\boldsymbol{x}_j\right).
\end{align}

In practice, the advantage of  QNG over standard gradients is quickly limited by the large computational cost of evaluating the QFI---which scales quadratically in the number of ansatz parameters $\boldsymbol{\theta}$. To circumvent this steep scaling, we use an efficient but approximate QNG-based optimizer that uses a simultaneous perturbation stochastic approximation (SPSA) \cite{1992Spall_SPSA} to compute both the underlying gradient and QFI at constant cost, i.e., QN-SPSA \cite{gacon2021simultaneous}.

SPSA gradient approximations are based on a perturbation of all system parameters in random directions
\begin{eqnarray}
\label{eq:gradient_approx}
 \nabla_{\boldsymbol{\theta}} L\left(\boldsymbol{\theta}\right) \approx \frac{L\left(\boldsymbol{\theta}+\epsilon \boldsymbol{\Delta}\right) - L\left(\boldsymbol{\theta}-\epsilon \boldsymbol{\Delta}\right)}{2\epsilon\boldsymbol{\Delta}}:= \boldsymbol{g}\left(\boldsymbol{\theta}\right),
\end{eqnarray}
where the vector $\boldsymbol{\Delta}$ represents the random directions and $0<\epsilon\ll 1$. 
Notably, the division by $\boldsymbol{\Delta}$ is element-wise.
If the random directions are sampled from a suitable distribution, e.g., uniformly drawn from $\{-1, 1\}$ for each parameter dimension, then the gradient approximation is known to represent an unbiased estimator \cite{1992Spall_SPSA}.
This concept can also be generalized to evaluate an SPSA approximation to the QFI with the following steps.
First, we rewrite the QFI as a Hessian
\begin{eqnarray}
\label{eq:qfi_hessian}
 \mathcal{F}^Q_{pq} = -2\frac{\partial^2\left\rvert\braket{\psi\left(\boldsymbol{\theta'}\right)\rvert\psi\left(\boldsymbol{\theta}\right)} \right\rvert^2}{\partial\theta_p\partial\theta_q}\Big\rvert_{\boldsymbol{\theta'}=\boldsymbol{\theta}} \:.
\end{eqnarray}
The equivalence of Eq.~\eqref{eq:qfi} and Eq.~\eqref{eq:qfi_hessian} is shown, e.g., in \cite{gacon2021simultaneous, Mari_2021Gradients}.
Next, we use a second-order SPSA scheme with two random direction vectors $\boldsymbol{\Delta}_1$ and $\boldsymbol{\Delta}_2$ \cite{Spall1997_SPSA} to approximate $\mathcal{F}^Q_{pq}$ stochastically such as
\begin{equation}
\label{eq:qfi_approx}
    \begin{aligned}
  \mathcal{\tilde{F}}^Q :=
 \frac{\boldsymbol{\Delta}_1\boldsymbol{\Delta}_2^{T}+\boldsymbol{\Delta}_2\boldsymbol{\Delta}_1^{T}}{-2\epsilon^2}
 &\Big(\left\rvert\braket{\psi\left(\boldsymbol{\theta}\right)\rvert\psi\left(\boldsymbol{\theta}+\epsilon \left(\boldsymbol{\Delta}_1 + \boldsymbol{\Delta}_2\right)\right)} \right\rvert^2
 - \left\rvert\braket{\psi\left(\boldsymbol{\theta}\right)\rvert\psi\left(\boldsymbol{\theta}+\epsilon \boldsymbol{\Delta}_1 \right)}\right\rvert|^2 - \\
 &\hspace{5mm}- \left\rvert\braket{\psi\left(\boldsymbol{\theta}\right)\rvert\psi\left(\boldsymbol{\theta}-\epsilon \left(\boldsymbol{\Delta}_1 -\boldsymbol{\Delta}_2\right)\right)} \right\rvert^2 +  \left\rvert\braket{\psi\left(\boldsymbol{\theta}\right)\rvert\psi\left(\boldsymbol{\theta}-\epsilon \boldsymbol{\Delta}_1 \right)} \right\rvert^2
 \Big)
 .
    \end{aligned}
\end{equation}
Finally, the SPSA approximations $\boldsymbol{g}\left(\boldsymbol{\theta}\right)$ and $\mathcal{\tilde{F}}^Q_{pq}$ are used to evaluate a QNG approximation as
\begin{align}
    \Big(\frac{1}{4}\mathcal{\tilde F}^Q\Big)^{-1} \boldsymbol{g}\left(\boldsymbol{\theta}\right),
\end{align}
which then enables the execution of a QN-SPSA optimization.
Since a single sample $\tilde{\mathcal{F}}^Q$ is generally not an accurate estimate of $\mathcal{F}^Q$, we typically use an average over a collection of independent samples. For more details we would like to refer the interested reader to \cite{gacon2021simultaneous}.

Lastly, the optimization scheme requires the initial state to have a non-zero overlap with the ground state, i.e., the optimal solution. A simple approach to suffice this criteria in our case is an initial parameterization $\boldsymbol{\theta}_{init}$, which allocates the same probability mass on every basis state, i.e.,
\begin{eqnarray}
\ket{\psi\left(\boldsymbol{\theta}_{\text{init}}\right)} = \frac{1}{\sqrt{2^n}}\sum\limits_{j=0}^{2^n-1}\ket{j}.
\end{eqnarray}


\subsection{Quantum Feature Selection}
\label{sec:VarQFS_application}

We now discuss the application of the variational quantum optimization as a wrapper method for feature selection---VarQFS. The following section presents the general methodology and the specific details on the numerical simulations and hardware experiments -- which verify the feasibility of our method -- are presented in  Sec.~\ref{sec:results}.

Suppose a training data set $X_{\text{train}}$ with $n$ input features and corresponding labels $y_{\text{train}}$. There are $2^n$ possible feature subsets labelled as $\set{\boldsymbol{x}_0, \ldots, \boldsymbol{x}_{2^{n-1}}}$. 
Furthermore, the basis states of the parameterized state $\ket{\psi\left(\boldsymbol{\theta}\right)}$ are identified with different combinations of feature subsets. The mapping from a basis state to a feature subset uses a binary approach: Every feature $l$ in the training data is related to a qubit $l$. If qubit $l$ is measured and the result corresponds to $\ket{1}$ ($\ket{0}$) then feature $l$ is included (excluded) from the feature subset.
During each optimization step, $N_\mathrm{shots}$ samples are drawn from the quantum state $\ket{\psi\left(\boldsymbol{\theta}\right)}$ and the measured states $\ket{m}$ are related to feature subsets $\boldsymbol{x}_m$.
Then, $\boldsymbol{x}_m$  and the training data $X_{\text{train}}$ are used to train a classification model $\mathcal{C}\left(\boldsymbol{x}_m, X_{\text{train}}\right)$. After training the model, a set of predicted labels can be extracted for the training data $y_m^{\text{pred}}$. These labels are further used to evaluate a performance score $\mathcal{S}\left(y_m^{\text{pred}}, y_{\text{train}}\right)$. 
In the following, we simplify the notation and write $\mathcal{S}_{\text{train}}\left(\boldsymbol{x}_m\right)$, i.e., the score corresponding to a selected feature subset evaluated on the training data, and $\mathcal{S}_{\text{test}}\left(\boldsymbol{x}_m\right)$, for the test data, respectively.

The goal is again to minimize the objective function with respect to the model parameters $\boldsymbol{\theta}$ such that the probability to sample feature subsets leading to the best objective values is maximized.
In practice, only a finite number of shots is drawn. Thus, the optimization is actually performed with the empirical loss function
\begin{align}
\label{eq:loss_VarQFS}
    \tilde{L}\left(\boldsymbol{\theta}\right) = \sum_{m=0}^{N_\mathrm{shots}-1} \tilde{p}_m\left(\boldsymbol{\theta}\right) \mathcal{S}_{\text{train}}\left(\boldsymbol{x}_m\right),
\end{align}
where empirical estimates of the sampling probabilities $\tilde{p}_m\left(\boldsymbol{\theta}\right)$ are evaluated as the normalized measurement frequencies of the basis states $\ket{m}$ after $N_{\text{shots}}$ samples.

\section{Results}
\label{sec:results}
This section presents VarQFS experiments for a publicly available credit risk data set \cite{Dua:2019_MLRepo}. First, we outline the details of the experiments and then we show the training results from experiments run with both,  a matrix product state (MPS) simulator, and the $\textit{ibmq\_montreal}$ quantum processing unit (QPU).
The presented proof-of-principle experiments show for both backends that variational quantum optimization has the potential to outperform traditional feature selection methods, i.e., RFE and RFECV, in certain aspects.
For the latter, Scikit-learn implementations \cite{scikit-learn2011} are used; for RFECV, we consider two different models: one trained using a log-loss score and the other using an accuracy score.

First,  the data set used for training and testing is presented in Sec.~\ref{sec:data}.
Next, the details of the VarQFS model and optimization are given in Sec.~\ref{sec:model_training} and the respective backends are discussed in Sec.~\ref{sec:backends}.
Then, we investigate the VarQFS training using both numerical simulation and a QPU for a reduced data set consisting of $20$ features (qubits) in Sec.~\ref{sec:reduced_data}. Finally, Sec.~\ref{sec:full_data} presents the results for training a VarQFS model on the full $59$-feature (qubit) problem with the MPS simulator. 
Notably, the goal of the MPS simulations is a proof-of-principle demonstration for the scalability of this approach.

\subsection{Data}
\label{sec:data}

This work uses a publicly available credit risk data set $X$ with corresponding labels $y$ which is made available by the UCI machine learning repository \cite{Dua:2019_MLRepo}. The data set includes $1,000$ data points which determine the credit-worthiness of a customer with respect to $20$ attributes. Some of these attributes are of a categorical nature. To incorporate these categorical attributes into our training, we employ one-hot encoding \cite{CategoricalDataforNeuralNetworksHancock20}. This expands the feature space of the data points to $59$ dimensional vectors.
This fact allows us to directly conclude that the individual features are not linearly independent. A list of all features is presented in Appendix~\ref{app:features_reduced}.

To train and validate the VarQFS model, the full data set is split into $50\%$ train and $50\%$ test data, $X_{\text{train}}$, $y_{\text{train}}$ and $X_{\text{test}}$, $y_{\text{test}}$, respectively.
Different ratios of test and train splits were attempted, along with different measures of multivariate distribution checks conditioned upon the response, through a process of parameter tuning.
To ensure that the generated train and test data sets are representative of each other,  we verify that the two data sets have comparable properties. To that end, the Pearson correlation coefficient \cite{PearsonCorrCoeff1895}---a linear correlation measure---is evaluated for every individual feature and the data instance label for the train and the test data. Next, we check that the respective Pearson correlation coefficients for train and test data vary at most by $\approx 10\%$.

\subsection{Model and Training}
\label{sec:model_training}
\begin{figure}[h!]
\begin{center}
\includegraphics[width=0.7\linewidth]{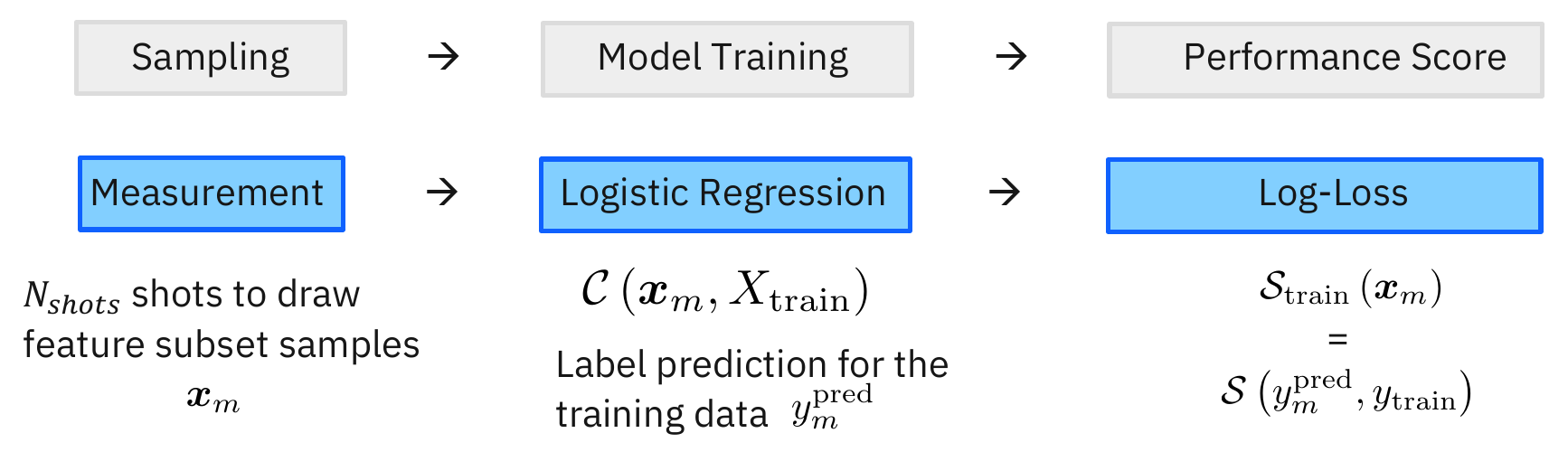}
\end{center}
        \captionsetup{singlelinecheck = false, format= hang, justification=centerlast, font=footnotesize, labelsep=space}
\caption{The diagram illustrates the workflow model for the VarQFS experiments presented in this section.}
\label{fig:qfs_workflow}
\end{figure}
We choose to implement baseline experiments using logistic regression as classifier $\mathcal{C}\left(\boldsymbol{x}_m, X_{\text{train}}\right)$ and log-loss as performance score $\mathcal{S}_{\text{train}}\left(\boldsymbol{x}_m\right)$ respectively $\mathcal{S}_{\text{test}}\left(\boldsymbol{x}_m\right)$, both provided by Scikit-learn \cite{scikit-learn2011}. 
The model training workflow is illustrated in detail in Fig.~\ref{fig:qfs_workflow}. Note that in our general formalism any arbitrary classifier and scoring function can be utilized as the model treats them as black boxes. 
In proof-of-principle experiments, we also tested VarQFS for other classifiers, e.g., the Random Forest and the Gradient Boosting classifier provided by Scikit-learn \cite{scikit-learn2011} as well as the LGBM classifier provided by LightGBM \cite{LightGBM} and observed comparable training behavior.
In our simulations, we (lazily) evaluate the scores throughout the optimization and cache the results. 
It should be noted that the scores are independent of the circuit parameters $\boldsymbol{\theta}$. 

\begin{figure}[h!]
\begin{center}
\includegraphics[width=0.7\linewidth]{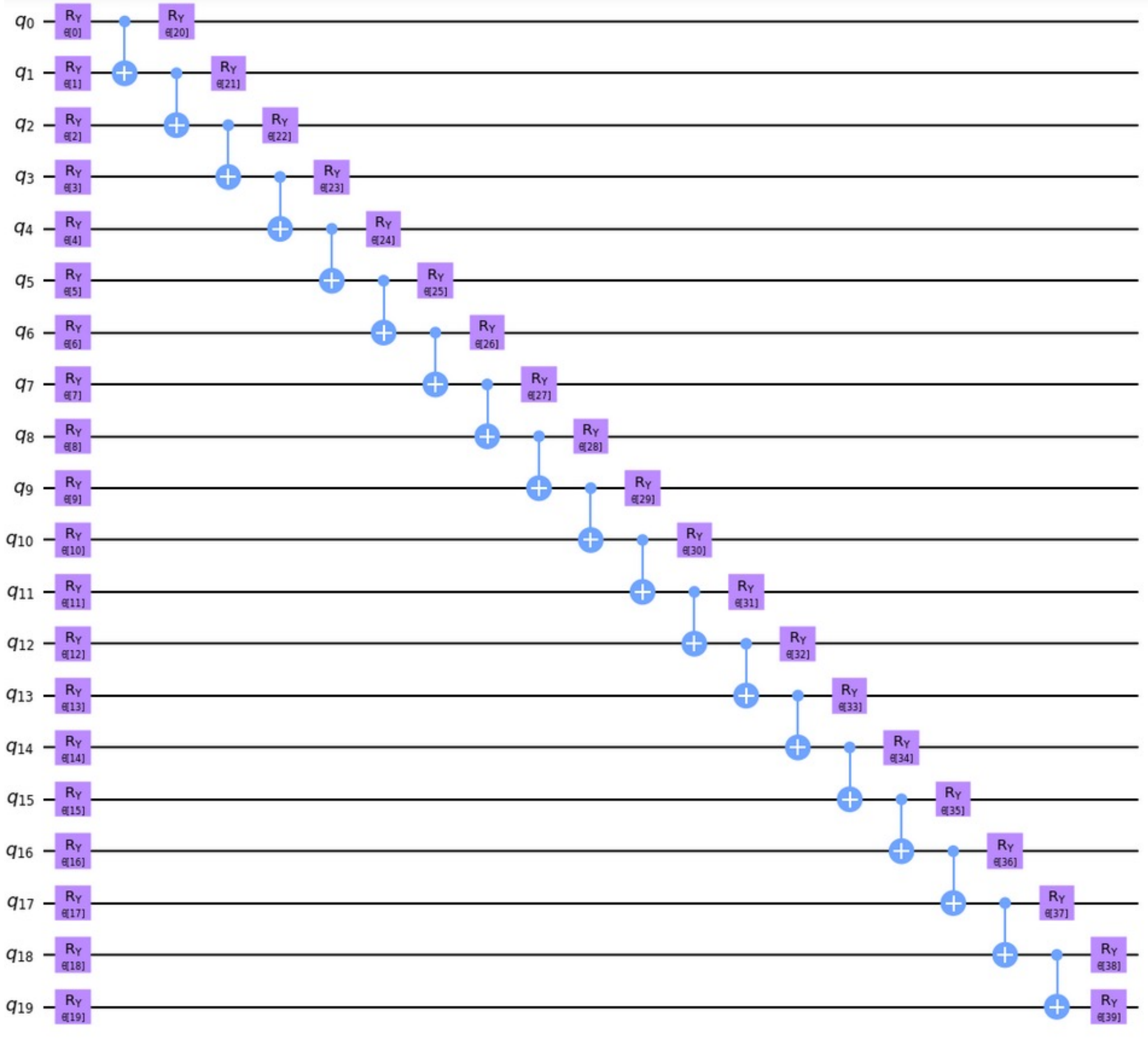}
\end{center}
\caption{This circuit shows a parameterized quantum circuit which acts as model ansatz. More specifically, this figure represents the `RealAmplitudes` ansatz for $20$ qubits/features with depth $1$ and `linear' entanglement.}
\label{fig:RAlin}
\end{figure}

Furthermore, we implement $\ket{\psi\left(\boldsymbol{\theta}\right)}$ using a parameterized quantum circuit ansatz class that applies solely real transformations onto the quantum state amplitudes. More specifically, we employ Qiskit's \textit{RealAmplitudes} ansatz with \textit{linear} entanglement which is illustrated in Fig.~\ref{fig:RAlin}.
The respective ans\"atze consist of single-qubit RY gates and of CNOT gates which are arranged in a linear fashion. Every ansatz consists of a layer of single-qubit gates and $d$ blocks of consecutive two-qubit and single-qubit gate layers. The number $d$ is often referred to as the circuit `depth'. In the experiments presented below $d$ is chosen as either $0$, $1$, or $2$ to enable accurate simulation with an MPS backend and sufficiently stable circuit execution on quantum hardware.
More specifically, this type of ansatz has a maximal bond dimension of $\chi \leq 2^d$ and can be represented exactly with an MPS simulator backend if $d$ is not too large (see Sec.~\ref{sec:backends} for more details).
Note that the QN-SPSA optimizer requires the evaluation of overlaps per Eq.~\eqref{eq:qfi_approx}, which is implemented via a compute-uncompute method: as the states are prepared by a known unitary, $\ket{\psi(\boldsymbol{\theta})} = U(\boldsymbol{\theta})\ket{0}^{\otimes n}$, the overlap of two states $|\braket{\psi(\boldsymbol{\theta}_1) | \psi(\boldsymbol{\theta}_2)}|^2$ can be computed by preparing the state
$U^\dagger(\boldsymbol{\theta}_2) U(\boldsymbol{\theta}_1)\ket{0}^{\otimes n}$ and evaluating the sampling probability of $\ket{0}^{\otimes n}$. This method to compute the overlap effectively doubles $d$ at fixed $n$. Thus, the required bond dimension is the \emph{square} of that for a single circuit, e.g., $\chi \leq (2^d)^2 = 4^d$ for the `linear' entanglement circuit.

In this case, ans\"atze with higher depths lead to more parameters. The accuracy of SPSA gradient and QFI approximations is better for systems with less parameters. Thus, we stabilize the training performance for ans\"atze with higher depths, by resampling the SPSA gradient and the SPSA QFI multiple times. The average of the gradient and  QFI samples are then used to compute the parameter updates with QN-SPSA.
In the following experiments, $5$ ($10$) samples define the updates if the ansatz has depth $1$ ($2$).

\subsection{Backends} \label{sec:backends}


To perform simulations at large qubit numbers, we employ an MPS simulator. 
The MPS formalism provides a classically efficient means of evaluating and analyzing moderately entangled quantum circuits~\cite{vidal_efficient_2003, schollwock_density-matrix_2011}. In our simulations, we use Qiskit \cite{qiskit} Aer's MPS simulator backend. The classical (space and time) computational cost of simulating circuits with MPS is, for a polynomial number of gates, only polynomial in the number of qubits $n$, but exponential in the amount of (bipartite) entanglement generated by the circuit. 
In this context, a natural measure of entanglement is given by $S_\chi = \log_2 \chi$, where $\chi$ is the maximal Schmidt rank over all possible bipartitions of the $n$ qubits~\cite{vidal_efficient_2003}. 
In the language of MPS, $\chi$ is equivalent to the (maximal) bond dimension of an \emph{exact} MPS representation of the state, i.e., with zero truncation error.
For our MPS simulations, we indeed consider only circuits with sufficiently limited entanglement and depth such that they can be represented exactly with relatively small bond dimension $\chi$.

The space complexity of the standard MPS method is $\mathcal{O}(n\chi^2)$ and the time complexity per evaluation of a single two-qubit gate is $\mathcal{O}(\chi^3)$. 
For MPS simulations of our generic loss function protocol spelled out above
requires that we must sample from $\ket{\psi(\boldsymbol{\theta})}$ to compute Eq.~\eqref{eq:loss}, thereby, as with hardware, incurring shot noise error. Sampling computational basis states of an MPS according to $p_x(\boldsymbol{\theta})$ is optimally done in time $\mathcal{O}(n\chi^2)$ per shot \cite{han_unsupervised_2018}.

\begin{figure}[h!]
\begin{center}
\includegraphics[width=0.7\linewidth]{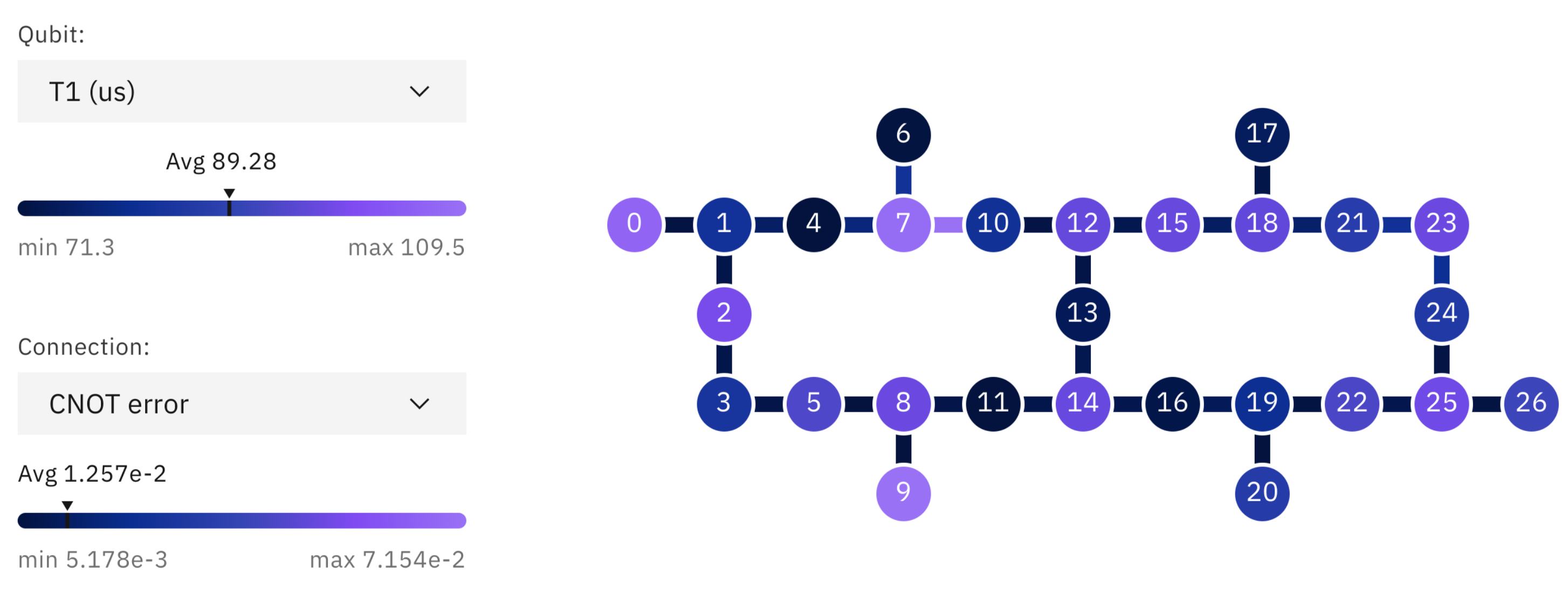}
\end{center}
        \captionsetup{singlelinecheck = false, format= hang, justification=centerlast, font=footnotesize, labelsep=space}
\caption{The topology diagram for $\textit{ibmq\_montreal}$ quantum processor. This diagram indicates the pairs of qubits that support two-qubit gate operations between them. This is also called the coupling map or connectivity. Qubits are represented as circles and the supported two-qubit gate operations are displayed as lines connecting the qubits. The coloring of the qubits indicates their respective $T_1$ times and the coloring of the connections between qubits indicates the errors induced by CNOT gates.}
\label{fig:topology}
\end{figure}

The chosen QPU is the $\textit{ibmq\_montreal}$ superconducting qubit processor with $27$ qubits. 
Fig.~\ref{fig:topology} depicts the topology diagram of this processor.
The diagram illustrates the $T_1$ times for each qubit as well as the CNOT gate error rates for each possible connection between qubits as depicted in the diagram.

\subsection{Reduced Data Set}
\label{sec:reduced_data}

As a first step, the application of VarQFS on $20$ features of the data, described in Sec.~\ref{sec:data}, is investigated.  These features are chosen as those $20$ features that are ranked as the most important ones according to an RFECV (log-loss) method that is applied to the full training data set. The respective features are listed in Appendix \ref{app:features_reduced}.
This setting is studied as it represents a non-trivial case that can still be evaluated with an exhaustive search method to find the feature subset with the best---in our case lowest---training score of $\mathcal{S}_{\text{train}}\left(\boldsymbol{x}_{\text{best}}\right) = 0.5351$. 
Notably, finding the subset which gives the best train score is the optimal solution of the objective function of the considered optimization problem.

In order to benchmark our VarQFS results, we evaluate the feature selection task with RFE and RFECV methods. 
RFECV is once run using a log-loss and once an accuracy scoring function $\mathcal{S}$. 
The final evaluation of all models is based on a log-loss score to ensure comparability with the VarQFS models.

\begin{figure}[!ht]
    \centering
    \begin{tikzpicture}
 \node at (3.5, 1) {\textbf{Reduced Data Set}};
\node[anchor=north west] at (-3.5, 0) {\includegraphics[width=0.45\textwidth]{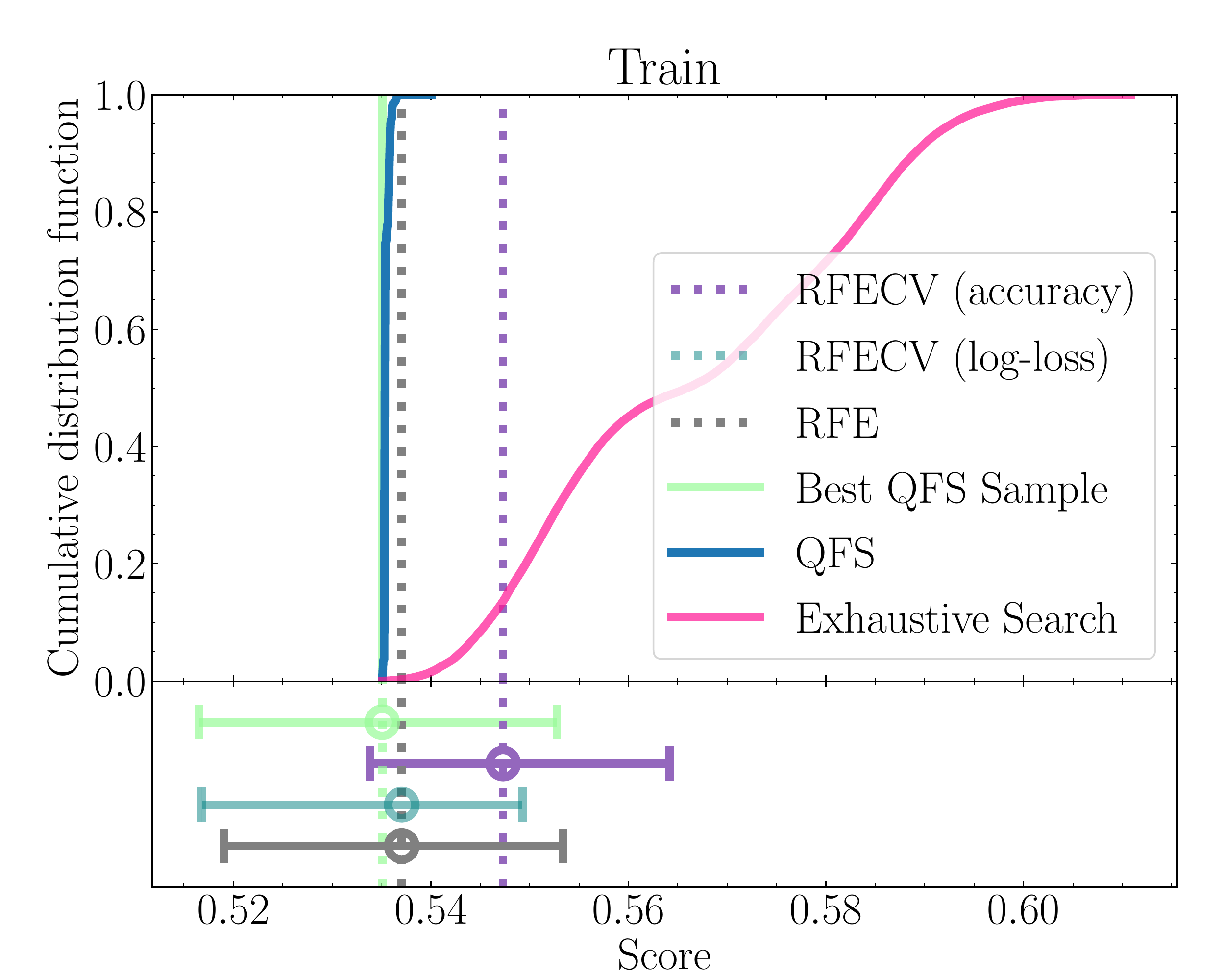}};
\node[anchor=north west] at (3.5, 0) {\includegraphics[width=0.45\textwidth]{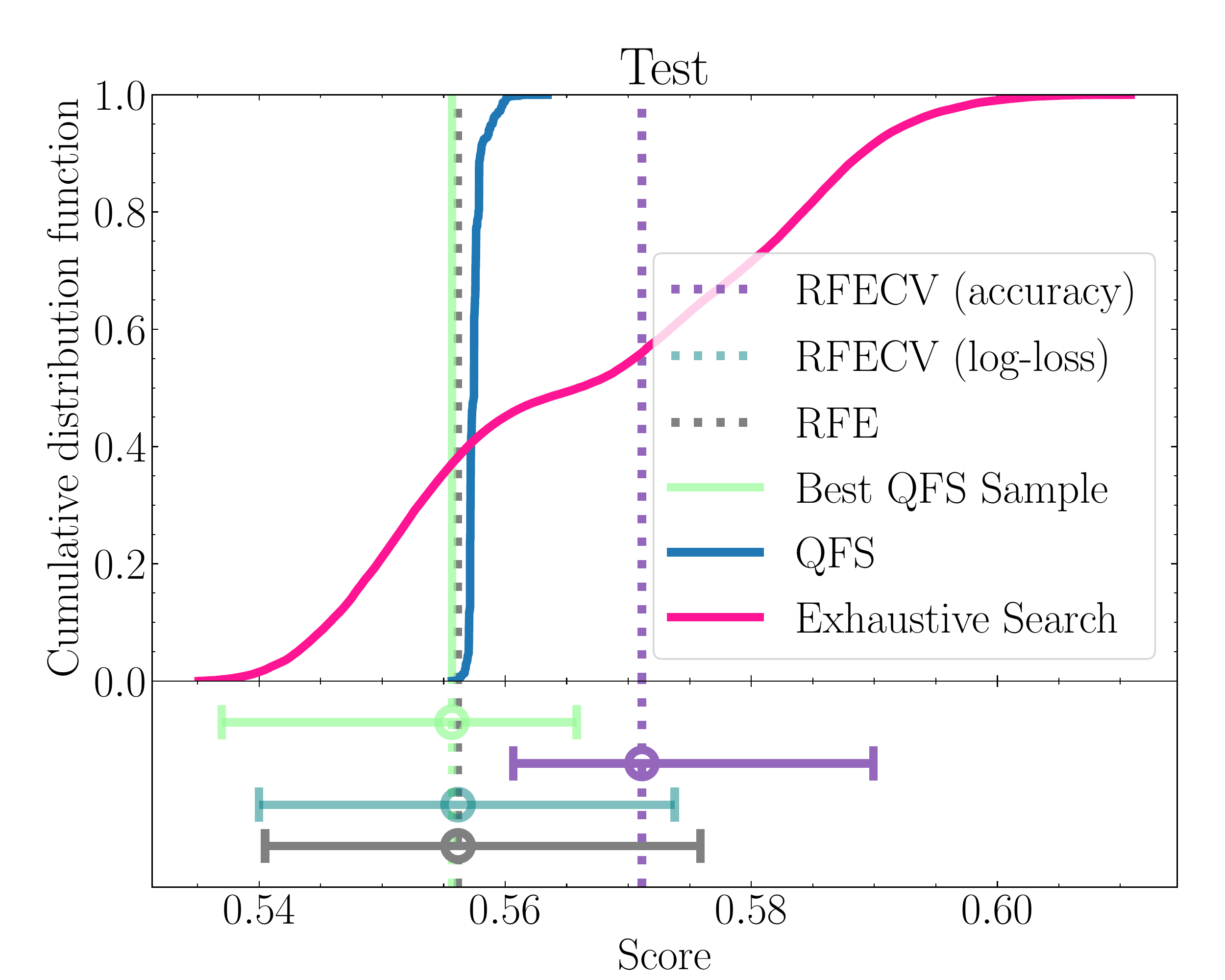}};
\node[anchor=north west] at (-3.5, -5.5) {\includegraphics[width=0.45\textwidth]{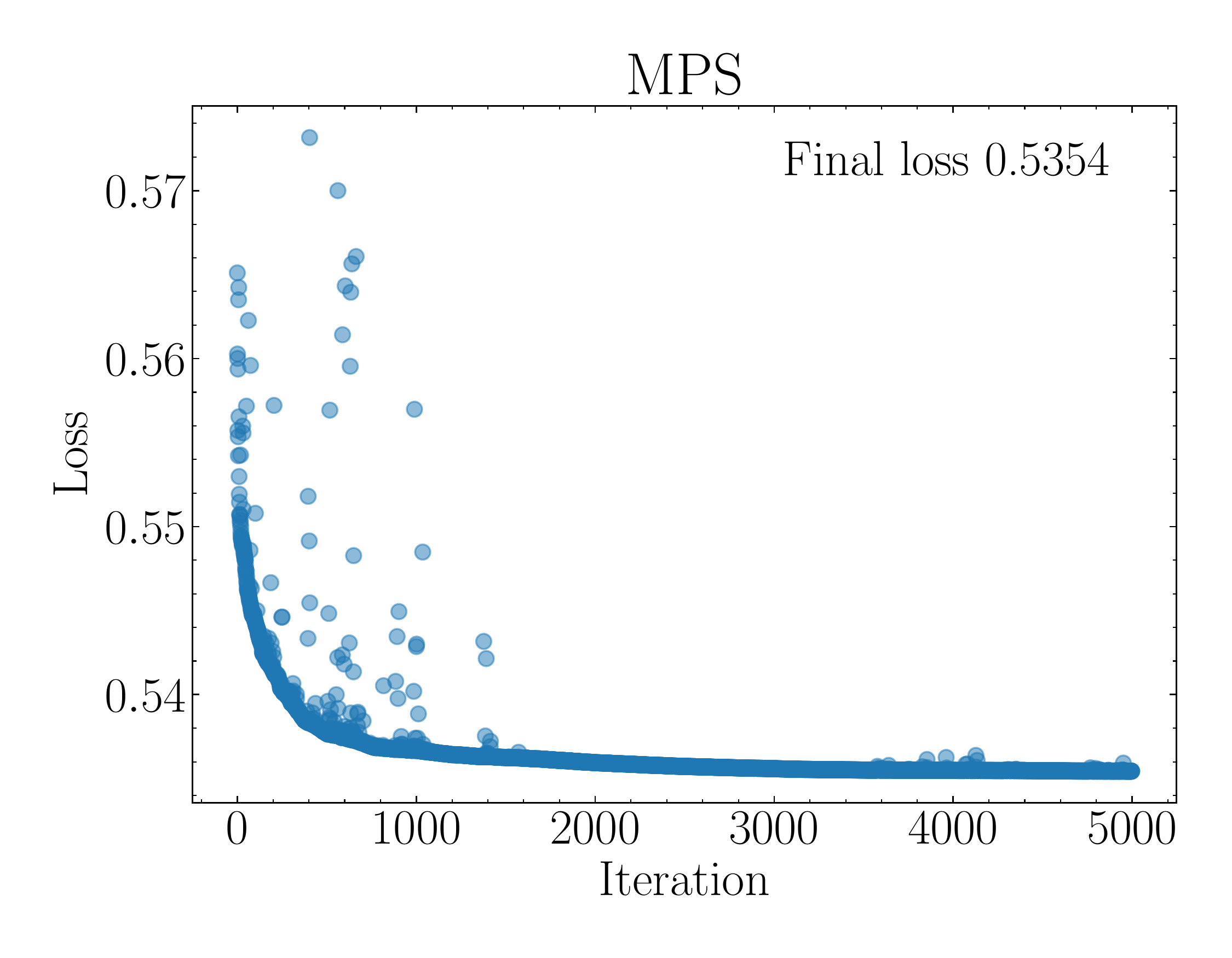}};
\node[anchor=north west] at (3.5, -5.5) {\includegraphics[width=0.45\textwidth]{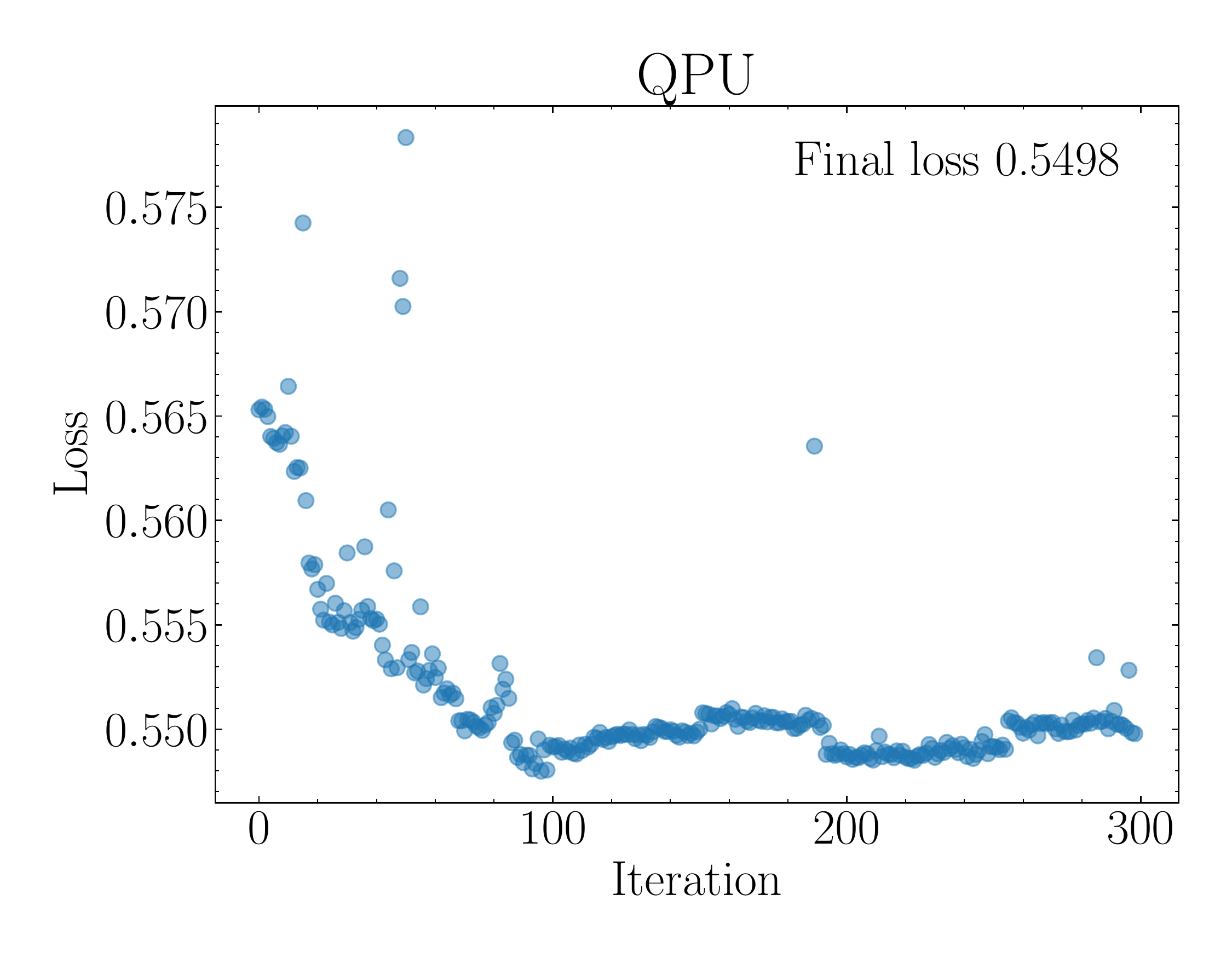}};
\node at (-2.5, -0.3) {(a)};
\node at (-2.5, -6) {(c)};
\node at (4.5, -0.3) {(b)};
\node at (4.5, -6) {(d)};
\end{tikzpicture}
        \captionsetup{singlelinecheck = false, format= hang, justification=centerlast, font=footnotesize, labelsep=space}
        \caption{
        (a) The figure shows the cumulative distribution function of the train scores with respect to the samples drawn from the trained MPS VarQFS model as well as the exhaustive search where all feature subsets are considered with equal sampling probability. Furthermore, the train scores of the best sampled subset, RFE and RFECV models are shown -- including the corresponding CIs. 
        (b) This figure is equivalent to (a) but is given for the test score. This shows that the features leading to the best train scores not necessarily lead to the best generalization performance.
        (c) The evolution of the loss function over $5,000$ iterations for the MPS VarQFS experiment is illustrated. 
        (d) The loss function progress for the QPU VarQFS experiment over $300$ iterations is presented.}
     \label{fig:20_qubits}
\end{figure}
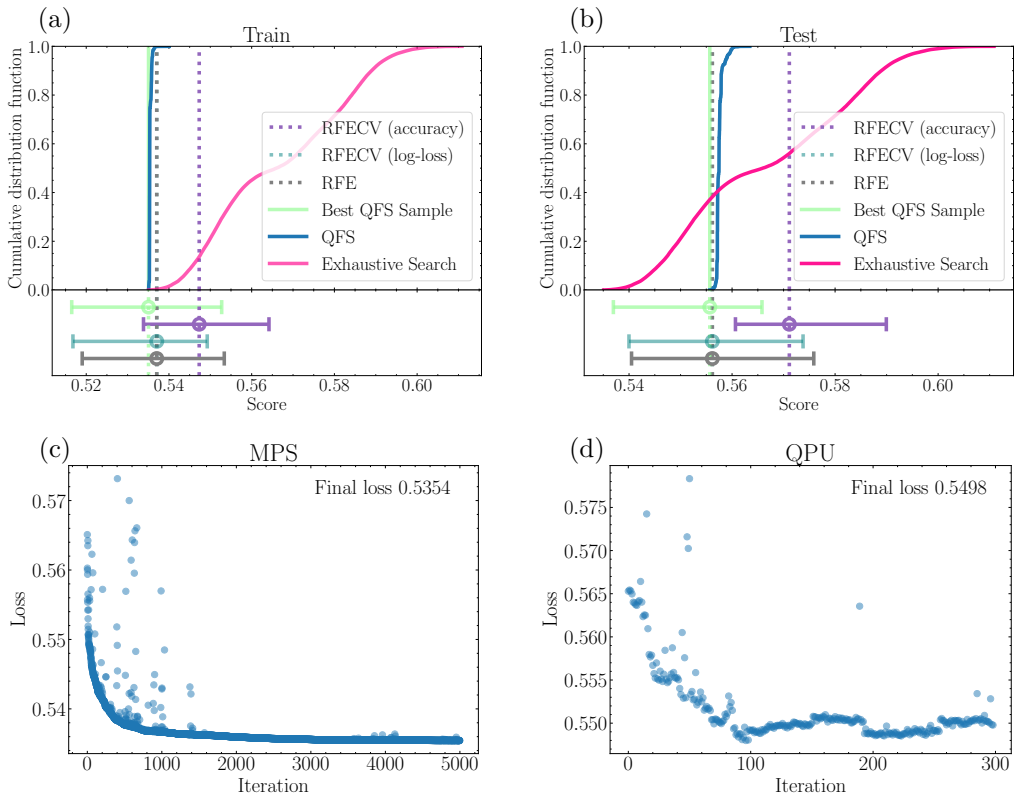

To run VarQFS with a depth $d=2$ ansatz,  we use the MPS simulator as backend and the QN-SPSA optimization for $5,000$ iterations where $N_\mathrm{shots}=10,000$ samples are drawn from $\ket{\psi(\boldsymbol{\theta})}$ in each iteration. The setting is trained for $10$ different random seeds to ensure that the presented results are statistically relevant. The final loss value of the training is on average $0.5357$ with a standard deviation of $\pm 0.0003$.
The example with the lowest loss after training is presented in Fig.~\ref{fig:20_qubits}.

After the training is finished, $10,000$ samples are drawn to analyse the sampling probabilities $p_m\left(\boldsymbol{\theta}\right)$ of feature subsets $\boldsymbol{x}_m$ of the trained quantum state. 
Fig.~\ref{fig:20_qubits}(a) and Fig.~\ref{fig:20_qubits}(b) present the cumulative distribution functions (CDFs) for the sampling probabilities $p_m\left(\boldsymbol{\theta}\right)$ with respect to the corresponding train $\mathcal{S}_{\text{train}}\left(\boldsymbol{x}_m\right)$ and test $\mathcal{S}_{\text{test}}\left(\boldsymbol{x}_m\right)$ scores. 
The figures also include the CDF scores corresponding to the exhaustive search where each feature subset has the same sampling probability.
Furthermore, the illustrations show the scores for the best samples and the respective scores found with the traditional RFE/RFECV methods as well as the confidence intervals (CIs).
Here and in the remainder of this work, all CIs are computed with confidence level $95\%$ using $50$ bootstraps with $70\%$ of the train/test data.

Fig.~\ref{fig:20_qubits}(a) illustrates that most of the probability mass of the state trained with VarQFS corresponds to feature subsets with lower train scores than any of the RFE/RFECV methods managed to find. The plot in Fig.~\ref{fig:20_qubits}(b) shows that VarQFS even manages to find a feature subset with a slightly better test score then RFE/RFECV. 
The figure also reveals that the feature subset with the smallest train score does not lead to the smallest test score.
Investigating the different feature subset samples drawn from the trained model can, thus, give important insights and help to make a practical decision.
The training evolution and its convergence is illustrated in Fig.~\ref{fig:20_qubits}(c) with respect to the loss function given in Eq.~\eqref{eq:loss_VarQFS}. The figure shows various outliers which stem from the stochastic nature of QN-SPSA. In order to avoid that these outliers interfere with the success of the optimization, the optimizer includes an additional check which discards any update that would increase the loss, up to a tolerance of twice the standard deviation of the loss function at the initial point \cite{Spall1997_SPSA}.

The VarQFS experiment carried out with \emph{ibmq\_montreal} employs again a depth $2$ ansatz---corresponding to $60$ one-qubit gates and $38$ two-qubit gates.
Unlike the simulated depth $2$ experiments, the hardware experiments do not employ additional resampling for the SPSA gradient/QFI due to the limited availability of quantum hardware resources.
The number of optimization iterations is also reduced to $300$. Furthermore, the number of samples drawn at each iteration is $8,000$; at the time of these experiments, the maximum number of samples that can be drawn by the QPU at a time is $8,192$. 

Although the loss function presented in Fig.~\ref{fig:20_qubits}(d) does not converge as smoothly as the MPS simulations, it still shows good convergence.
The illustrated outliers are in this case partially due to the stochastic nature of QN-SPSA and partially due to the physical noise of the quantum backend. Again, the optimization scheme only accepts parameter updates which improve the loss function.

A comparison of the results from all feature selection methods described above for the $20$ feature problem is presented in Tab.~\ref{tbl:20_feature_experiments}. 
The table includes the train/test scores and corresponding CIs for the features selected with RFE and RFECV.
Furthermore, the best train/test scores sampled in the analysis step, and corresponding CIs are given for the VarQFS experiments executed with the MPS simulator as well as the QPU. 
We would like to point out that the MPS results are averaged for all $10$ executions with different random seeds.

\begin{table}[!htb]
\captionsetup{singlelinecheck = false, format= hang, justification=raggedright, font=footnotesize, labelsep=space}
\centering{
\scriptsize{
\begin{tabular}{c|c|c|c|c}
Method & Train Score & Test Score & Train CI & Test CI \\
 \hline
RFECV (acc) & $0.5473$ & $0.5711$ & $[0.5344, 0.5633]$  & $[0.5527, 0.5845]$\\
RFECV (log) & $0.5371$ & $0.5562$ & $[0.5184, 0.5551]$ & $[0.5391, 0.5731]$ \\
RFE &  $0.5371$ & $0.5562$ & $[0.5229, 0.5525]$ & $[0.5368, 0.5737]$ \\
MPS & $\boldsymbol{0.5351 \pm 1.16\cdot 10^{-5}}$ & $0.5558 \pm 0.0003$ & $[0.5172 \pm 0.0028, 0.5517 \pm 0.0022]$ & $[0.5372 \pm 0.0035, 0.5722 \pm 0.0027]$ \\
QPU &  $\boldsymbol{0.5351}$ & $\boldsymbol{0.5542}$ & $[0.5162, 0.5542]$ & $[0.5386, 0.5715]$ \\
\end{tabular}
}}
\caption{The table summarizes the results for the experiments conducted with $20$ features. For the RFECV (accuracy), RFECV (log-loss) and RFE feature selection methods the train scores, test scores and corresponding CI bounds are shown.
The VarQFS experiments are presented with respect to the best train score, the best test score and corresponding CIs.
Notably, the results from the MPS simulations are given in terms of the average and standard deviation for the optimizations with $10$ different random seeds.}
\label{tbl:20_feature_experiments}
\end{table}
The table reveals that while the simulated as well as QPU VarQFS experiments manage to find feature subsets leading to the optimal train score, the classical methods do not.
Furthermore, the data reveals that although the loss function for the QPU VarQFS experiment does not converge as nicely as in the simulated experiment, the sampled feature subsets lead to a similarly good train and even better test score than the MPS VarQFS result average.
Besides good training results, we are in practice interested in finding a feature subset which also leads to good generalizability expressed by a small test score $\mathcal{S}_{\text{test}}\left(\boldsymbol{x}_m\right)$. 
The VarQFS methods manage to find (on average) better test scores than the classical methods.
Moreover, we would like to point out that it is the small amount of training data which is (most likely) the culprit of the relatively large CIs.

We now discuss the relative performance of the quantum and classical methods in more detail. To this end, we conduct a statistical analysis comparing the train and test scores of the trained quantum models with RFECV (accuracy), RFECV (log-loss) and RFE feature selection methods.
We examined paired scores on $1000$ data subsets with $70\%$ of the training and test data and evaluate the percentage of subsets on which the VarQFS algorithm outperforms the classical methods. 
More specifically, we first find the $10$ feature subsets sampled from the trained VarQFS algorithm that lead to the smallest train scores on the full training data. Next, we evaluate the percentage of data sets where VarQFS gives better train and test scores than RFECV (accuracy), RFECV (log-loss) and RFE. Finally, these percentages are averaged over the $10$ feature subsets.
The results are presented in Tab.~\ref{tbl:stat_analysis20}.

\begin{table}[!htbp]
\captionsetup{singlelinecheck = false, format=hang, justification=raggedright, font=footnotesize, labelsep=space}
\scriptsize{
\begin{tabular}{c|c|c|c|c|c}
Method & Compared Model & Avg. Train Data & Max. Train Data  & Avg. Test Data & Max. Test Data \\
\hline
MPS & RFECV (acc) & $100\%$& $100\%$ & $100\%$ & $100\%$ \\ 
QPU & RFECV (acc)& $100\%$& $-$ & $100\%$ &	$-$   \\
MPS & RFECV (log)& $100\%$& $100\%$ & $100\%$ & $100\%$ \\ 
QPU & RFECV (log)& $99.8\%$& $-$ & $96.4\%$ &	$-$   \\
MPS & RFE & $100\%$& $100\%$ & $100\%$ & $100\%$ \\ 
QPU & RFE & $99.7\%$& $-$ & $96.3\%$ &	$-$   \\
\end{tabular}
}
\caption{The statistical analysis examines paired scores on $1000$ data subsets consisting of $70\%$ of the original train and test data for the reduced data set.
The table shows the percentage of instances where VarQFS leads to better scores on the train and test data compared to RFECV (accuracy), RFECV (log-loss) and RFE.
For the VarQFS algorithm implemented with the MPS simulator, we evaluate the average over the $10$ different seeds used for training as well as maximum percentage for the various seeds. 
Since the experiments with the QPU are only executed on one seed, the presented value only represents an average over the $10$ feature subsets for this instance.}
\label{tbl:stat_analysis20}
\end{table}
We find that the VarQFS algorithm is able to outperform the classical methods in almost all of these instances.
It can be seen that the average performance for the best seed across the training data for all three classical methods is quite high, with the quantum algorithm achieving a lower log-loss score on between $89\%$ and $100\%$ of the data sets. This indicates that the quantum algorithm has the potential to outperform classical methodologies.

\begin{table}[!htb]
\captionsetup{singlelinecheck = false, format= hang, justification=raggedright, font=footnotesize, labelsep=space}
\centering{
\small{
\begin{tabular}{c|c|c|c}
Backend & Iteration ($s$) & Gradient ($s$) & QFI ($s$)\\
\hline
MPS Simulator & $11.1294$ & $0.0008$ & $0.0064$\\
QPU & $818.5777$ & $\approx 24$ (w.o.~queuing)& $\approx 48$ (w.o.~queuing)\\
\end{tabular}
}}
\caption{This table presents the time needed to run a single iteration (without resampling), the SPSA gradient, as well as, SPSA QFI evaluation with the MPS simulator and the actual hardware.
The time measured for one iteration with the QPU corresponds comes from an exemplary run with priority access to the \textit{ibmq\_montreal} backend.
Notably, the SPSA gradient/QFI times correspond to the time needed for backend compilation, instrument loading and execution on the QPU. The time it takes to queue for circuit execution is not included.}
\label{tbl:times_20qubits}
\end{table}

Finally, Tab.~\ref{tbl:times_20qubits} shows that the time required to run the VarQFS algorithm on an actual quantum backend is still significantly larger than the efficient MPS simulator. 

\subsection{Full Data Set}
\label{sec:full_data}

Next, we investigate the performance of VarQFS on the full credit risk data set from Sec.~\ref{sec:data}.
Since the number of features in this setting is too large, we can no longer conduct an exhaustive search for comparison.
Nevertheless, traditional feature selection methods---in our case RFE/RFECV---can be used as benchmark for the VarQFS performance. The respective train/test scores and corresponding CIs are presented in Tab.~\ref{tbl:log-loss59}.

To train VarQFS on the given data set $X_{\text{train}}$, $10$ different random seeds are chosen for ans\"atze with depth $0$, $1$ and $2$. Each training employs $5,000$ iteration steps with QN-SPSA drawing $10,000$ samples in every step.
The convergence is on average slightly better for $d=0$ than for higher depths. More explicitly, $d=0$ results on average in loss values of $0.4389$ with standard deviation $\pm 0.0009$, while $d=1$ ($d=2$) results on average in $0.4510\:(0.4432)$ with standard deviation $\pm 0.0144\:(0.0024)$. 
This may be due to the approximations of the QN-SPSA optimizer.
Since the ans\"atze with higher depths have more parameters, the QN-SPSA approximations are more likely to deviate from the actual gradient, and QFI.
While the QN-SPSA implementation already employs additional resamplings for the ans\"atze with higher depths, we expect that an increase in the number of resamplings would lead to further improvement in the convergence behavior.
Nevertheless, higher depths are on average better in terms of the best test score samples. 

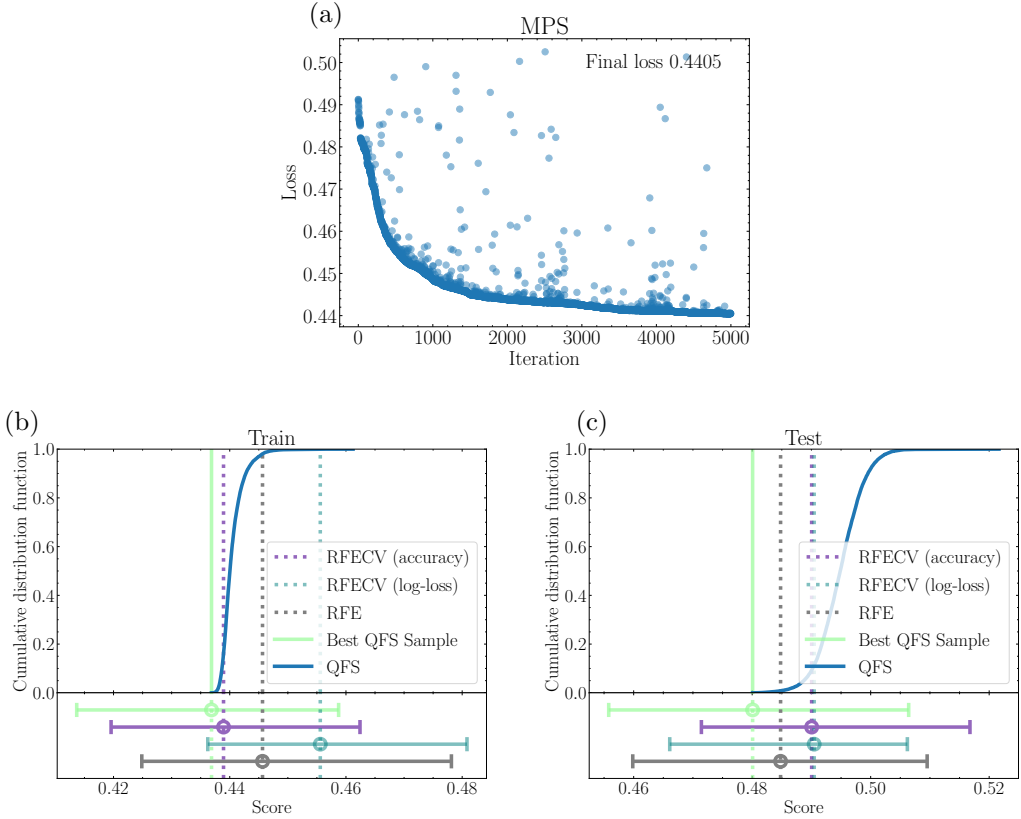
\begin{figure}[!ht]
    \centering
    \begin{tikzpicture}
 \node at (3.5, 1) {\textbf{Full Data Set - Simulation}};
\node[anchor=north west] at (0, 0) {\includegraphics[width=0.45\textwidth]{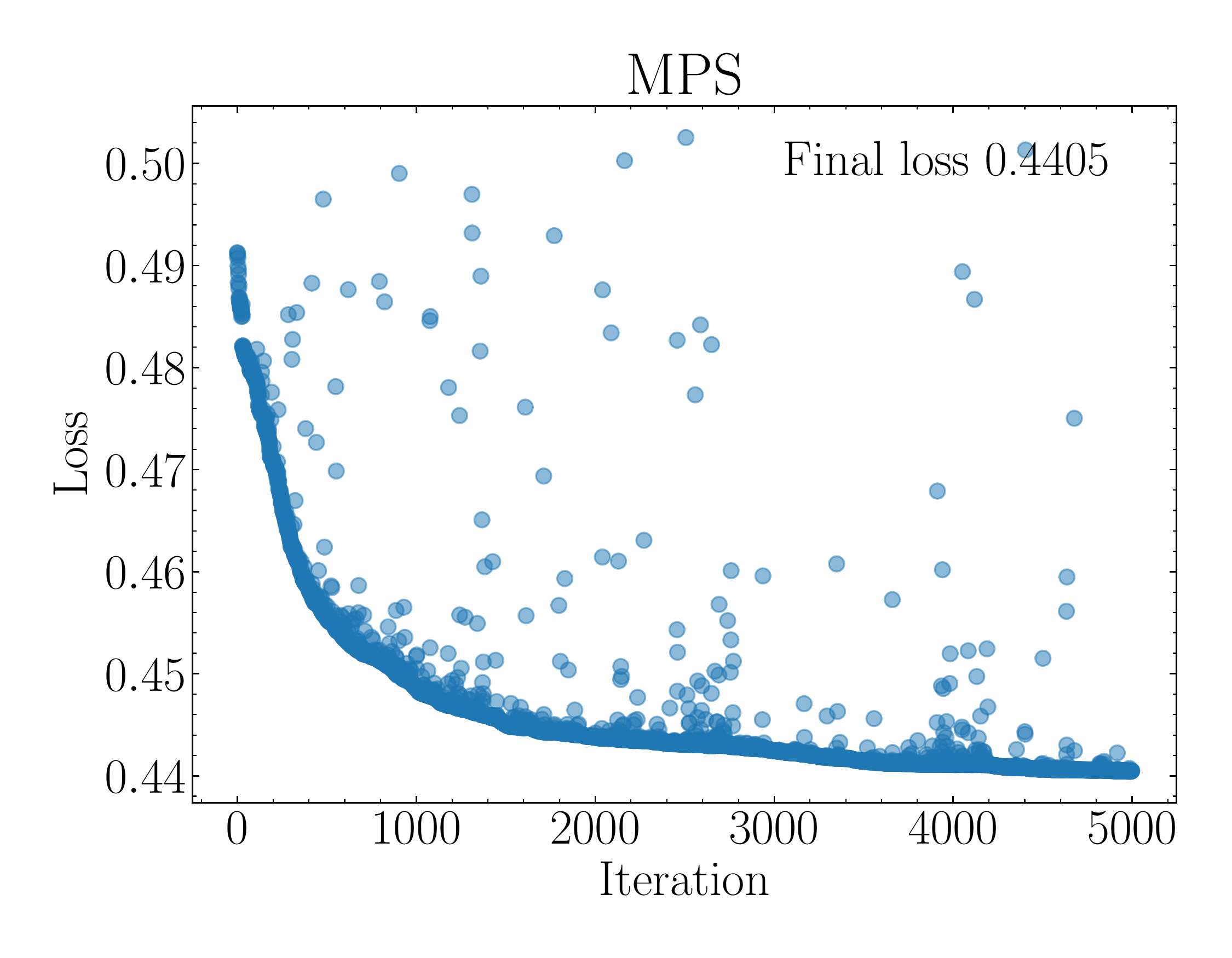}};
\node[anchor=north west] at (-3.5, -5.5) {\includegraphics[width=0.45\textwidth]{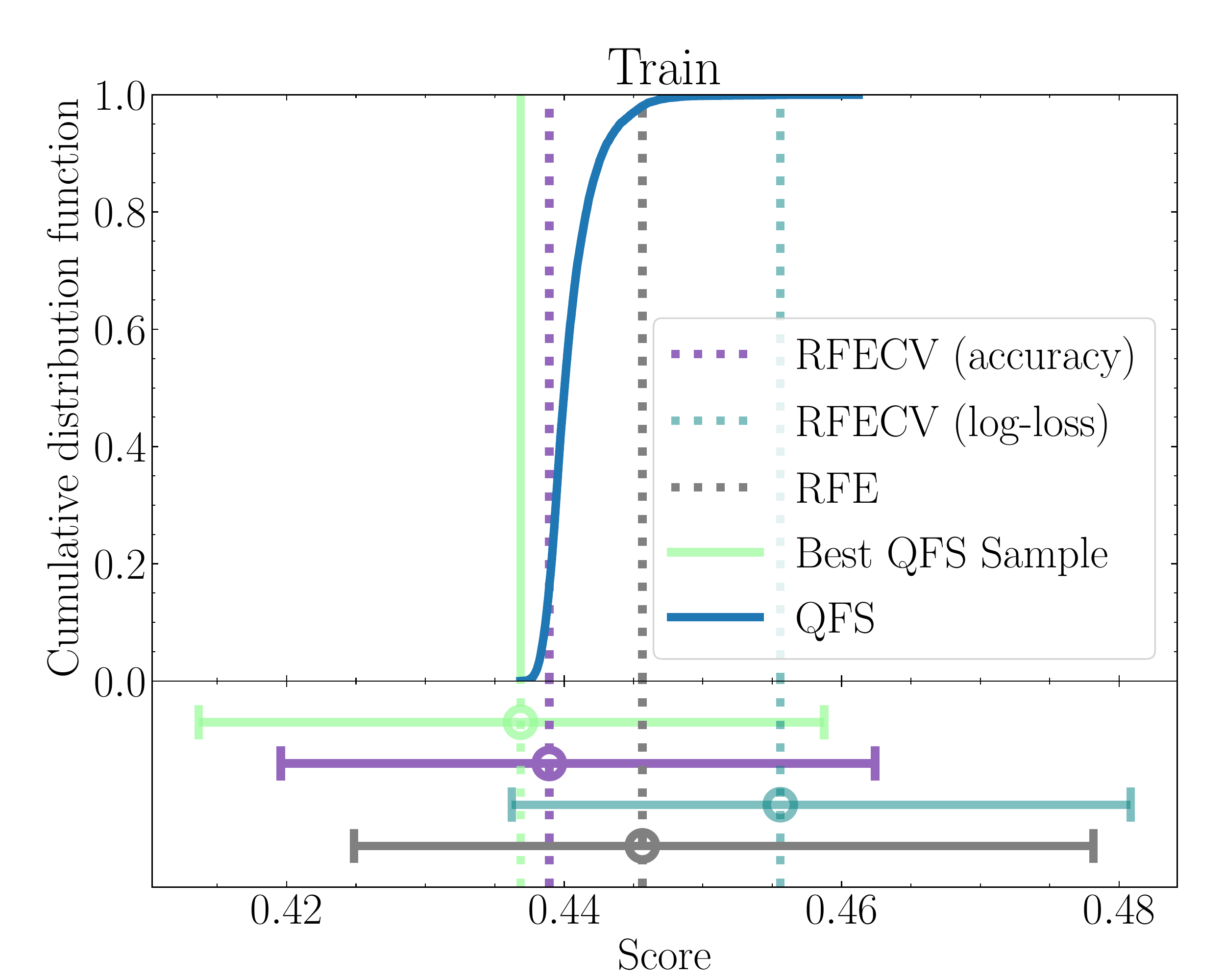}};
\node[anchor=north west] at (3.5, -5.5) {\includegraphics[width=0.45\textwidth]{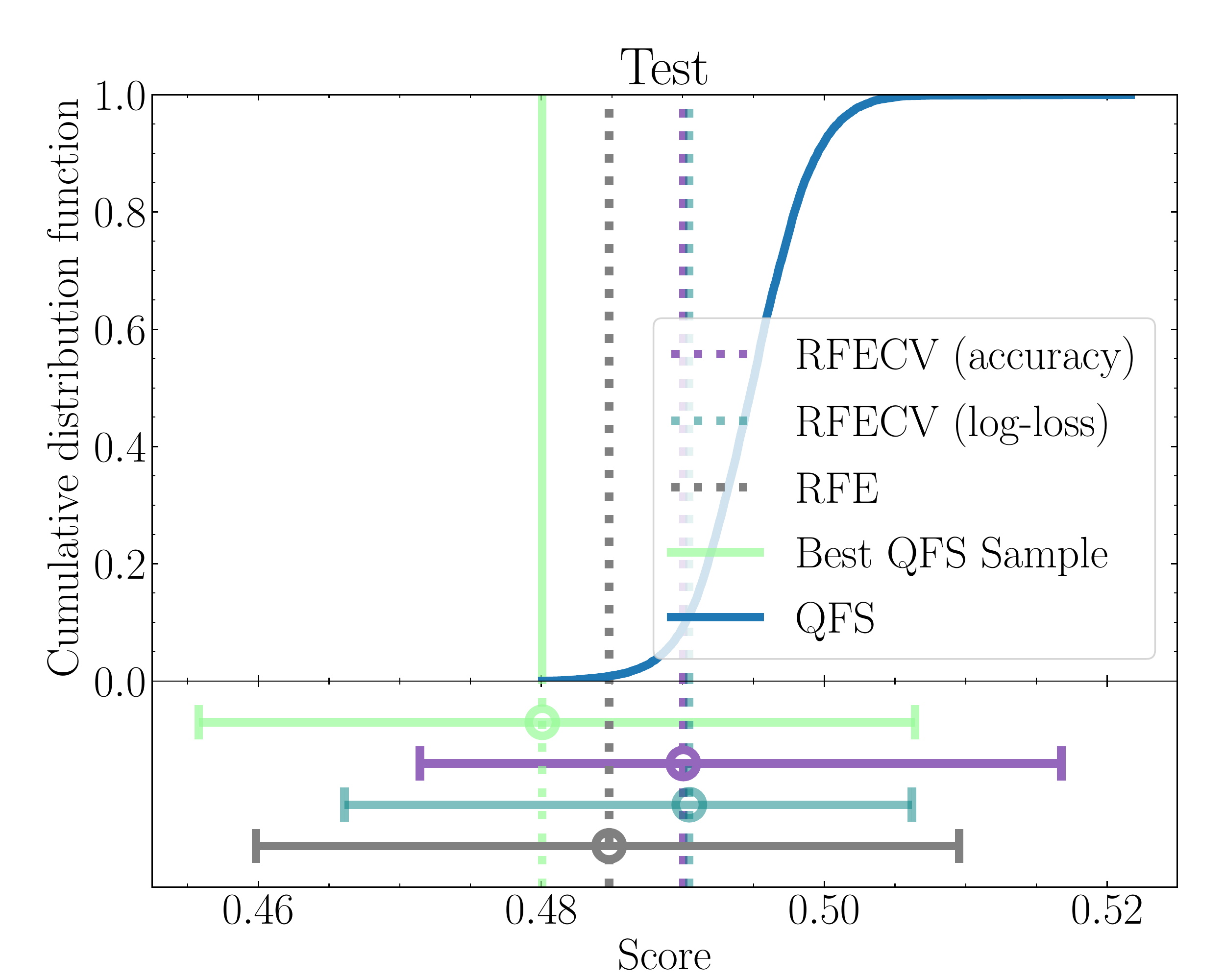}};
\node at (1, -0.4) {(a)};
\node at (-3, -5.8) {(b)};
\node at (4.5, -5.8) {(c)};
\end{tikzpicture}
        \captionsetup{singlelinecheck = false, format= hang, justification=centerlast, font=footnotesize, labelsep=space}
        \caption{(a) The evolution of the loss function over $5,000$ iterations is illustrated. (b) The figure shows the cumulative distribution function of the train scores with respect to the samples drawn from the trained VarQFS model. Furthermore, the train scores of the best sampled subset, RFE and RFECV models are shown -- including the corresponding CIs. (c) This figure is equivalent to (b) but is given for the test scores.}
     \label{fig:59_qubits_sim}
\end{figure}

Fig.~\ref{fig:59_qubits_sim} presents the  training with depth $2$ that lead to the best loss score. The loss function in Fig.~\ref{fig:59_qubits_sim}(a) shows good convergence behavior despite the occurrence of various outliers during the optimization.
Fig.~\ref{fig:59_qubits_sim}(b) and Fig.~\ref{fig:59_qubits_sim}(c) present the CDF for the train and test scores corresponding to the feature subset probability distribution evaluated with $10,000$ samples drawn from the trained VarQFS model. 
Furthermore, the figures illustrate the best sample drawn from VarQFS and the scores evaluated with RFE/RFECV as well as the corresponding CIs.
The graphs not only show that the VarQFS algorithm manages to find feature subsets that are better than the ones found by RFE/RFECV, but also, that significant probability mass lies on feature subsets whose training scores are better than the RFECV (log-loss) and RFE solutions.

Next, we extract the final objective values and the best train and test scores as well as the corresponding CIs from all $10$ optimization runs for each depth $0$, $1$ and $2$. We report their averages and standard deviations in Tab.~\ref{tbl:log-loss59}, whereby the train and test scores are evaluated using $10,000$ samples.

\begin{table}[!htb]
\captionsetup{singlelinecheck = false, format=hang, justification=raggedright, font=footnotesize, labelsep=space}
\centering{
\scriptsize{
\begin{tabular}{c|c|c|c|c}
Method & Train Score & Test Score & Train CI & Test CI \\
 \hline
RFECV (acc) & $0.4389$ & $0.4900$ & $[0.4193, 0.4651]$ & $[0.4643, 0.5132]$ \\
RFECV (log) & $0.4556$ & $0.4905$ &  $[0.4288, 0.4772]$ & $[0.4658, 0.5099]$\\
RFE  & $0.4456$ & $0.4848$ & $[0.4191, 0.4638]$ &  $[0.4597, 0.5188]$ \\
MPS $d=0$ & $\boldsymbol{0.4386\pm 0.0034}$ &  $0.4804\pm 0.0028$ & $[0.4164 \pm 0.0050, 0.4621\pm 0.0044]$ & $[0.4537 \pm 0.0043, 0.5033 \pm 0.0059]$ \\
MPS $d=1$  & $0.4388 \pm 0.0034$ &  $0.4776 \pm 0.0038$ & $[0.4150 \pm 0.0055, 0.4596 \pm 0.0043]$ & $[0.4545 \pm 0.0050, 0.5026 \pm 0.0063]$\\
MPS $d=2$ & $0.4401 \pm 0.0043$ &$\boldsymbol{0.4759 \pm 0.0019}$& $[0.4182\pm 0.0068, 0.4620\pm 0.0026] $ & $[0.4514 \pm 0.0035, 0.4982\pm 0.0048]$\\
\end{tabular}
}}
\caption{The table shows the train scores, test scores and corresponding CI bounds for the RFECV (accuracy), RFECV (log-loss) and RFE feature selection methods for the logistic regression model with $59$ features.
The train scores for $d=0$ are on average marginally better than higher depth results.
This may again be related to the approximation error of QN-SPSA mentioned before.
Moreover, the mean value and standard deviation for the experiments conducted with $10$ random seeds are presented for each depth $0$, $1$ and $2$ with respect to the best train/test scores of each experiment and corresponding CI.}
\label{tbl:log-loss59}
\end{table}
Furthermore, the mean values for the best sampled train and test scores are better than the classical RFE/RFECV methods for all depths.
The analysis also reveals that the best train score $\mathcal{S}_{\text{train}}\left(\boldsymbol{x}_m\right)=0.4386$ and test score $\mathcal{S}_{\text{test}}\left(\boldsymbol{x}_m\right)=0.4759$ is found for depth $0$ and $2$.

Throughout the various experiments good trainability is observed.
In Appendix.~\ref{sec:trainability}, we discuss the occurrence possibilities of barren plateaus \cite{McClean_2018BarrenPlateaus} in VarQFS. Furthermore, we present empirical evidence that the VarQFS model investigated in this paper is unlikely to be affected by this problem.

Finally, we employ the same statistical analysis as in Sec.~\ref{sec:reduced_data} to investigate the statistical significance of the VarQFS performance in comparison to RFECV (accuracy), RFECV (log-loss) and RFE. Tab.~\ref{tbl:stat_analysis_59} reflects the percentage of instances on which VarQFS outperforms the classical algorithms. 

\begin{table}[!ht]
\captionsetup{singlelinecheck = false, format=hang, justification=raggedright, font=footnotesize, labelsep=space}
\scriptsize{
\begin{tabular}{c|c|c|c|c|c}
Method & Compared Model & Avg. Train Data & Max. Train Data  & Avg. Test Data & Max. Test Data \\
\hline
MPS $d=0$ & RFECV (acc) & $67.3\%$ & $90.45\%$& $42.1\%$& $65.4\%$ \\ 
MPS $d=1$ & RFECV (acc)& $62.0\%$ & $89.9\%$ & $33.5\%$& $61.1\%$ \\ 
MPS $d=2$ & RFECV (acc)& $50.0\%$ & $89.9\%$ & $35.0\%$& $53.5\%$ \\ 
MPS $d=0$ & RFECV (log)& $94.4\%$ & $100\%$& $48.0\%$& $64.8\%$ \\ 
MPS $d=1$ & RFECV (log)& $97.1\%$ & $100\%$ & $43.3\%$& $57.4\%$ \\ 
MPS $d=2$ & RFECV (log)& $95.3\%$ & $100\%$& $42.7\%$& $53.9\%$ \\ 
MPS $d=0$ & RFE & $84.3\%$ & $99.8\%$& $17.0\%$& $84.0\%$ \\ 
MPS $d=1$ & RFE &$85.5\%$ & $99.7\%$& $14.7\%$& $71.2\%$ \\ 
MPS $d=2$ & RFE & $69.3\%$ & $99.7\%$& $10.7\%$& $15.2\%$ \\ 
\end{tabular}
}
\caption{The statistical analysis examines paired scores on $1000$ data subsets consisting of $70\%$ of the original train and test data for the full data set.
The table shows the percentage of instances where VarQFS lead to better scores on the train and test data compared to RFECV (accuracy), RFECV (log-loss) and RFE.
For the VarQFS algorithm implemented with the MPS simulator, we evaluate the average over the $10$ different seeds as well as maximum percentage for the various seeds. }
\label{tbl:stat_analysis_59}
\end{table}
It can be observed that there is a strong seed dependence in the performance characteristics of these experiments. 
We hypothesise that this is related to the training issues mentioned above. More specifically, we expect that given the current number of QN-SPSA resamples the VarQFS method does not converge to its optimal point for all seeds given the full data set and that the statistical test reflects this with lower fractions of advantageous VarQFS feature selection runs.
While the statistical significance of the presented results is not as strong as for the reduced data set, one can still see that VarQFS manages in every setting at least for one seed to outperform the classical methods on more than $\approx 90\%$ of the training data instances. Even for the test data, there exists---except for one instance---one seed where VarQFS outperforms RFECV (accuracy), RFECV (log-loss) and RFE in more than half of the cases.

\section{Discussion, Conclusion, and Outlook}
\label{sec:conclusion}

This work presents a variational quantum optimization algorithm for unconstrained black box binary optimization.
With a relation of the optimization task to a ground state problem and by employing a solver motivated through quantum imaginary time evolution we provide strong theoretical justification for promising convergence behavior.
We have demonstrated the potential of this approach to compete with, and even outperform, established classical methods on the example of feature selection for a real-world data set using MPS simulation as well as a real QPU.

The proof-of-principle results presented here---from the simulation as well as the actual quantum hardware---do in fact illustrate the potential of VarQFS to find better feature subsets than traditional methods.
We would like to highlight that the $\textit{ibmq\_montreal}$ based VarQFS model manages to outperform traditional methods even without using error mitigation \cite{ErrorMitigationTemme17, ErrorMitPiveteau21}. 
The empirical study for the full data set reveal that using higher depth ans\"atze---which ultimately leads to more entanglement in the system---can improve the generalization of the model, i.e., help to find on average better test scores.
While the results show good convergence for the VarQFS training, we expect that further studies of QN-SPSA and its hyperparameter settings can help to further improve the training, particularly for higher depth ans\"atze. 
Furthermore, it would be an interesting research question to investigate the VarQITE error bounds presented in \cite{zoufal2021error} in the context of a stochastic QN-SPSA implementation which could help to get an improved understanding of the accuracy of the presented methods.

One particularly promising property of this method in the context of feature selection is that all features and their correlations are considered jointly. Many classical methods on the other hand consider only up to pair-wise correlations and are, thus, prone to missing important information about the underlying data.
In this context, it would be interesting to compare VarQFS to classical feature selection methods, e.g., based on probabilistic models \cite{ProbabilisticFeatureSelectionSaito18}, that also consider all variable correlations jointly. In fact, it should be possible to generalize the ideas of, for example, Refs.~\cite{han_unsupervised_2018, stoudenmire_supervised_2017, novikov_exponential_2017} and use MPS themselves (or some other tensor network structure) as a probabilistic model for difficult combinatorial optimization tasks such as feature selection.

We would also like to highlight that the $59$ feature (qubit) MPS experiments represent relatively large simulations of a full variational quantum algorithm, especially given the necessity of sampling shots in calculating not only the loss function but also the overlaps in the definition of the approximate QFI. These results lay the foundation for future investigations of VarQFS experiments at larger scale with actual quantum hardware.
This, in addition to the classical complexity of evaluating the black box scoring function for a given sample, make these simulations particularly complicated and expensive.
While this type of simulation has rarely been explored for variational quantum algorithms, we believe that it has a lot of potential in this context, especially for understanding the role that increasing entanglement may play in such algorithms.
While we have focused on representing the problem exactly with MPS, an interesting question that remains to be studied in variational quantum algorithms more broadly is whether or not \emph{approximate} MPS simulations of training would produce barren plateaus induced by finite truncation error and its accompanying state infidelity---an effect which might be interpreted as a type of noise \cite{zhou_what_2020, wang_noise-induced_2021}.

Furthermore, it can be expected that future quantum hardware will enable to study this algorithm in the context of larger systems and ans\"atze with more entanglement. 
In particular, we would then not have to rely on classical simulators such as the MPS simulator but could investigate the results of scaled experiments with non-trivial ansatz classes.
The impact of entanglement in the model ansatz is an important point that is open for future research. While our empirical results revealed that ans\"atze with higher depth can improve certain aspects of the trained model, it remains to be studied whether this is due to a certain structure in the entanglement or the larger parameter space. Here, we have focused on a low-depth ansatz with RY gates and a simple `linear' entanglement structure of CNOT gates, due to its compatibility with both the employed QPU and MPS simulator (in the sense of requiring only 1D nearest-neighbor connectivity), but understanding the appropriateness of other ansatz circuit choices---including perhaps those built out of quantum (e.g., isometric) tensor networks \cite{huggins_towards_2019, foss-feig_holographic_2021, haghshenas_variational_2021, slattery_quantum_2021, maccormack_simulating_2021}---is an important open question. In general, it would be interesting to study if preliminary statistical (classical) analysis could be employed to find a suitable entanglement structure.

Future work could also study further variations of the optimization settings.
For example, the method can easily be adapted to ensure that the final feature subset contains $m$ features by either using a Hamming-weight preserving ansatz or by introducing penalty terms to the loss function.
Moreover, one could compare VarQFS to a larger variety of classical state-of-the-art feature selection methods by conducting further empirical studies. 
Other potential improvements in the settings, such as training methods based on cross-validation techniques, may also be investigated.

Finally, we would like to conclude by pointing out that while this work presents the potential of variational quantum optimization for feature selection, it remains a question for future research to study this approach on other combinatorial optimization problem instances or even generalizations of the objective function type.

\vspace{3mm}

\noindent\textbf{Acknowledgments.}
We thank Hubert Le Van Gong, Siew Hoon Lim and Vidyut Naware for continuous constructive and inspiring discussions.
We would also like to thank Hedayat Alghassi for his help in executing simulations, and Merav Aharoni and Hiroshi Horii for discussions and work on the Aer MPS simulator. 

\vspace{3mm}

\noindent\textbf{Code Availability.}
The code for the implementation of the classical parts of the algorithm can be found under Scikit-learn \cite{scikit-learn2011}. Furthermore, the code for the implementation of the variational quantum circuit as well as the QN-SPSA optimizer used for the presented experiments are publicly available in Qiskit \cite{qiskit}. Lastly, the data set used in the presented experiments is made available by the UCI machine learning repository \cite{Dua:2019_MLRepo}.

\appendix
\section{Features Data Set}
\label{app:features_reduced}
This section provides a list of the features included in the reduced data set and the full data set from the German credit risk data set from the UCI Machine Learning Repository \cite{Dua:2019_MLRepo}.
The following list presents all features of the given data set.

\begin{itemize}
    \item Status of existing checking account
    \begin{itemize}
        \item A11: $x<0$
        \item A12: $0\leq x<200$
        \item A13: $x\geq 200$
        \item A14: no checking account
    \end{itemize}
    \item Duration in month
    
    \item Credit History
    \begin{itemize}
        \item A30: no credits taken/ all credits paid back duly
        \item A31: all credits at this bank paid back duly
        \item A32: existing credits paid back duly till now
        \item A33: delay in paying off in the past
        \item A34: critical account/ other credits existing (not at this bank)
    \end{itemize}
    
    \item Purpose
    \begin{itemize}
        \item A40: car (new)
        \item A41: car (used)
        \item A42: furniture/equipment
        \item A43: radio/television
        \item A44: domestic appliances
        \item A45: repairs
        \item A46: education
        \item A47: (vacation - does not exist?)
        \item A48: retraining
        \item A49: business
        \item A410: others
    \end{itemize}
    \item Credit Amount
    \item Savings account/bonds
    \begin{itemize}
        \item A61: $x<100$
        \item A62: $100 \leq x < 500$
        \item A63: $500 \leq x < 1000$
        \item A64: $x \geq 1000$
        \item A65: unknown/ no savings account
    \end{itemize}
    \item Present employment since
    \begin{itemize}
        \item A71: unemployed
        \item A72: $x < 1 $year
        \item A73: $1 \leq x < 4 $ years
        \item A74: $4 \leq x < 7$ years
        \item A75: $x \geq 7$ years
    \end{itemize}

    \item Installment rate in percentage of disposable income
    \item Personal status and sex
    \begin{itemize}
        \item A91: male: divorced/separated
        \item A92: female: divorced/separated/married
        \item A93: male: single
        \item A94: male: married/widowed
        \item A95: female: single
    \end{itemize}
    \item Other debtors / guarantors
    \begin{itemize}
        \item A101: none
        \item A102: co-applicant
        \item A103: guarantor
    \end{itemize}
    \item Present residence since
    \item Property
    \begin{itemize}
        \item A121: real estate
        \item A122: if not A121: building society savings agreement/ life insurance
        \item A123: if not A121/A122: car or other, not in attribute 6
        \item A124: unknown / no property
    \end{itemize}
    \item Age in years
    \item Other installment plans
    \begin{itemize}
        \item A141: bank
        \item A142: stores
        \item A143: none
    \end{itemize}
    \item Housing
    \begin{itemize}
        \item A151: rent
        \item A152: own
        \item A153: for free
    \end{itemize}
    \item Number of existing credits at this bank
    \item Job
    \begin{itemize}
        \item A171: unemployed/ unskilled - non-resident
        \item A172: unskilled - resident
        \item A173: skilled employee / official
        \item A174: management/ self-employed/
highly qualified employee/ officer
    \end{itemize}
    \item Number of people being liable to provide maintenance for
    \item Telephone
    \begin{itemize}
        \item A191: none
    \item A192: yes, registered under the customers name
    \end{itemize}
    \item foreign worker
    \begin{itemize}
        \item A201: yes
        \item A202: no 
    \end{itemize}
\end{itemize}

The following list presents the features are ranked as the $20$ most important from the full data set by RFECV (log-loss).

\begin{itemize}
    \item Existing Checking Account Status A11
    \item Saving Account Or Bonds A65
     \item Present Employment Since A71
     \item Present Employment Since A72
      \item Present Employment Since A74
      \item  Marital Status And Gender A93
      \item Other Debtors Or Guarantors A102
      \item Other Debtors Or Guarantors A103
      \item
      Property A124
      \item Other Installment Plans A141
      \item Other Installment Plans A143
      \item Housing Status A151
      \item Housing Status A153
      \item Housing Status A152
      \item
      Present Employment Since A75
      \item Marital Status And Gender A94
      \item Credit History A32
      \item Current Residence Duration 
      \item Employment Status A174
      \item Other Installment Plans A142
\end{itemize}

\section{Barren Plateaus}
\label{sec:trainability}

In this section, we discuss the scalability of the suggested VarQFS algorithm with respect to the potential occurrence of exponentially vanishing gradients -- also called \textit{barren plateaus} \cite{McClean_2018BarrenPlateaus}.
In particular, we give empirical evidence that the given setting of model and training data is unlikely to be subject to cost function induced barren plateaus \cite{CerezoCostFunctDependentBarrenPlats21}.

Barren plateaus may occur due to various reasons.
Firstly, ans\"atze which are very expressive, such that they represent (approximate) $2$-designs \cite{Hayashi_2005tdesigns}, can lead to a landscape which is exponentially flat in the number of qubits $n$ \cite{McClean_2018BarrenPlateaus}. This may, e.g., be circumvented by informed initialization strategies \cite{Grant_2019_InitializationStrategyForAddressingBarrenPlateaus, Rudolph2022_QuantumCircuits_TensorNetworks} or the design of problem-dependent ans\"atze \cite{Meyer2022_SummetryQML}.
Secondly, exhaustive entanglement in an ansatz---in combination with partial traces on parts of the state---may also lead to barren plateaus \cite{Wiebe2020Barren, sharma2020trainability, Holmes_2021BPs_scramblers}. Entanglement-induced barren plateaus may be avoided by limiting the entanglement in the chosen model ansatz \cite{Patti_2021EntanglementDevisedBPs}.
Thirdly, exponentially vanishing gradients can be due to the local noise of current quantum hardware \cite{wang_noise-induced_2021}. At the moment, hardware noise may be reduced using error mitigation \cite{ErrorMitigationTemme17, ErrorMitPiveteau21}. Fault-tolerant quantum computers would eradicate this issue.
Lastly, barren plateaus can occur due to cost functions that depend on \textit{global observables}, i.e., observables which require all $n$ qubits to be measured at the same time \cite{CerezoCostFunctDependentBarrenPlats21}.
The first three culprits of barren plateau may be handled independently of the problem at hand, e.g., by using the measures suggested above.
But the last type is an intrinsic property of the algorithm's objective function.

In this case, training data is an important factor when it comes to the potential impact of cost function dependent barren plateaus. More explicitly, we see from Eq.~\ref{eq:diag_h} that the image of $f$ for $\set{\boldsymbol{x_0}, \ldots, \boldsymbol{x_{2^n-1}}}$ defines the observable of the given optimization problem. A-priori we do not know whether the image of $f$ induces a global observable or local observable---an observable that can be evaluated by measuring at most $k\ll n$ qubits together and does not lead to gradients that vanish exponentially in the system size.

In the remainder of this section, we give numerical evidence that the presented VarQFS problem is unlikely to suffer from this problem for the given credit risk data set and may, thus, be scalable in our setting.
To that end, we numerically verify that the observable induced by the image of $f$ for the given credit risk data can be (approximately) represented by a local Hamiltonian.
More specifically, it is shown that a linear program can fit a quadratic binary unconstrained optimization (QUBO) problem \cite{QUBO14Kochenberger}, which corresponds to a $2-$local Hamiltonian, to the objective values of our problem for the given credit risk data set for $n=\set{5, 10, 15, 20}$ qubits.

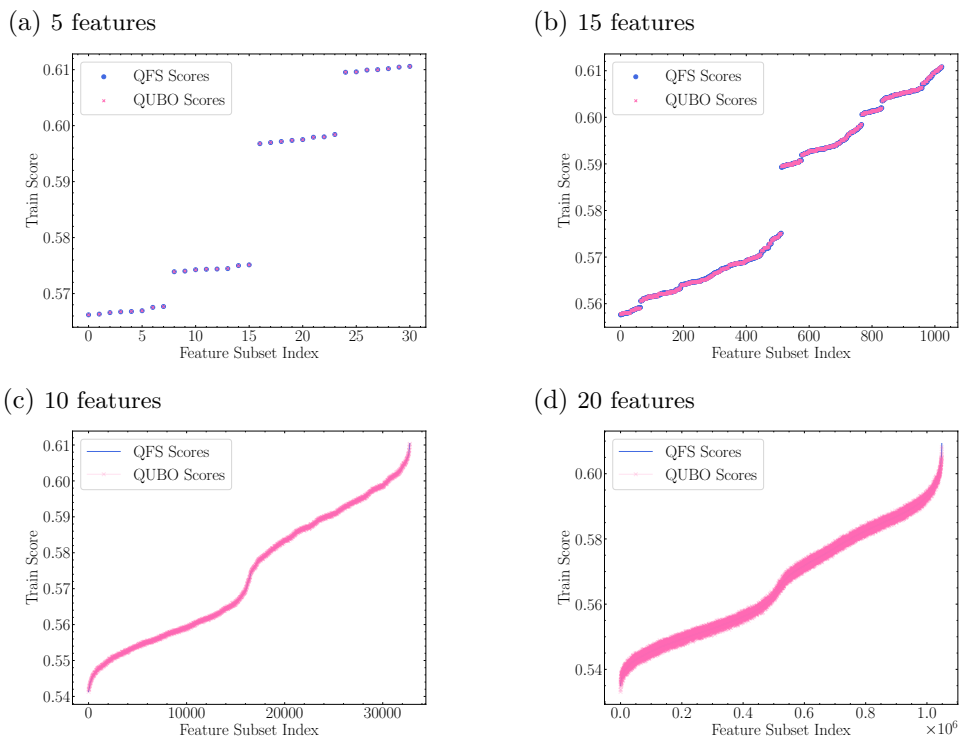
\begin{figure}[!ht]
    \centering
    \begin{tikzpicture}
\node[anchor=north west] at (-6, 0) {\includegraphics[width=0.4\textwidth]{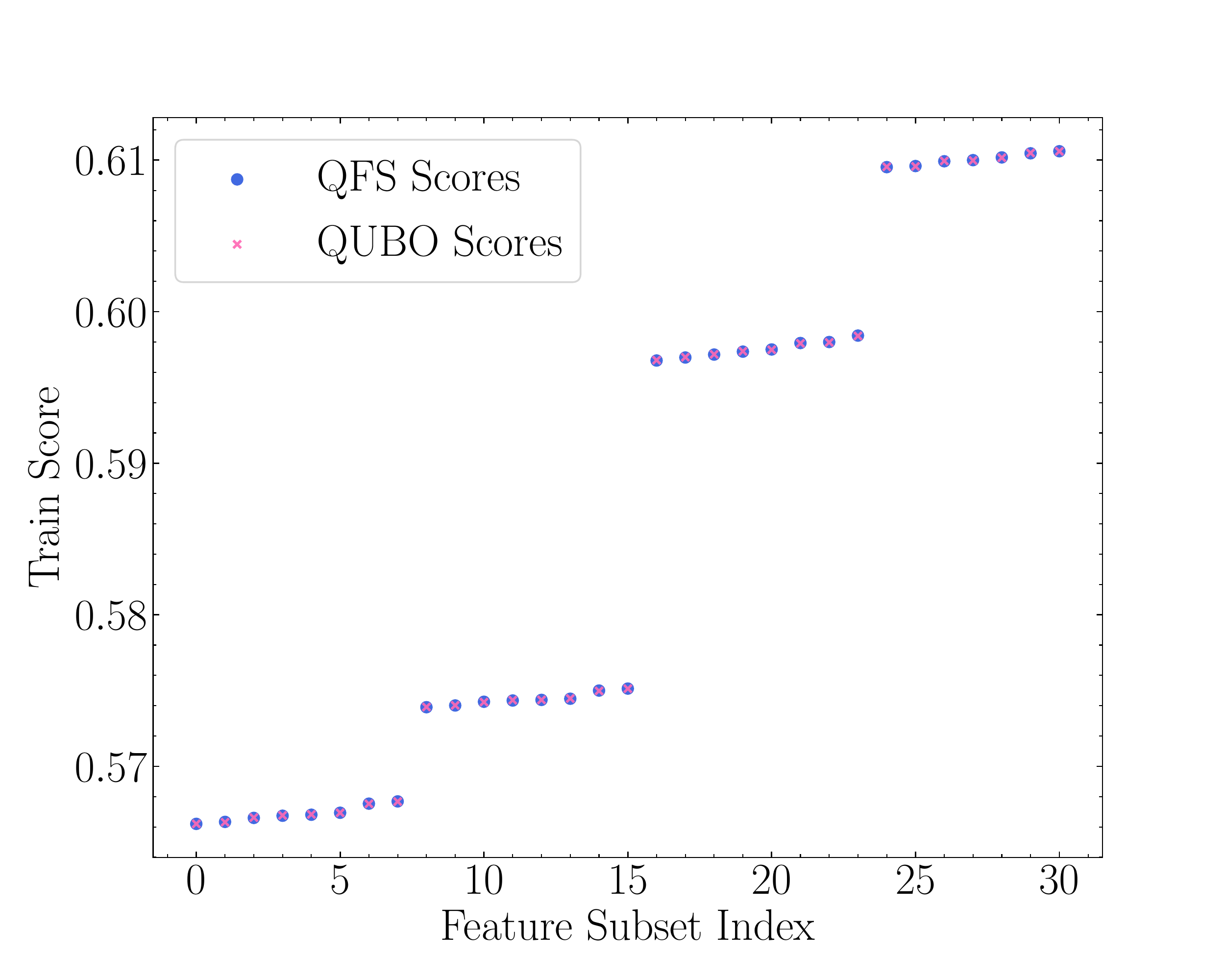}};
\node[anchor=north west] at (-6, -5) {\includegraphics[width=0.4\textwidth]{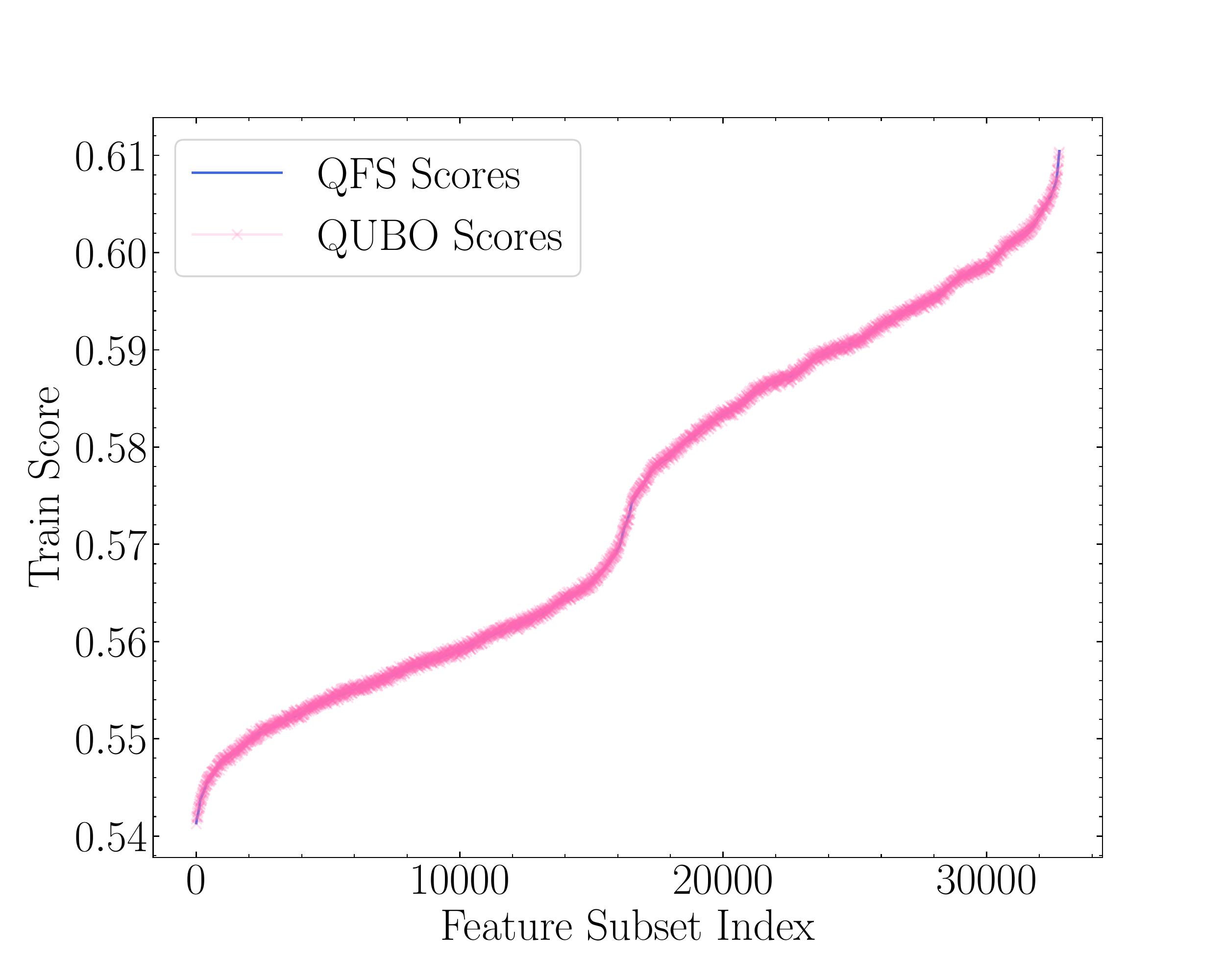}};
\node[anchor=north west] at (1, 0) {\includegraphics[width=0.4\textwidth]{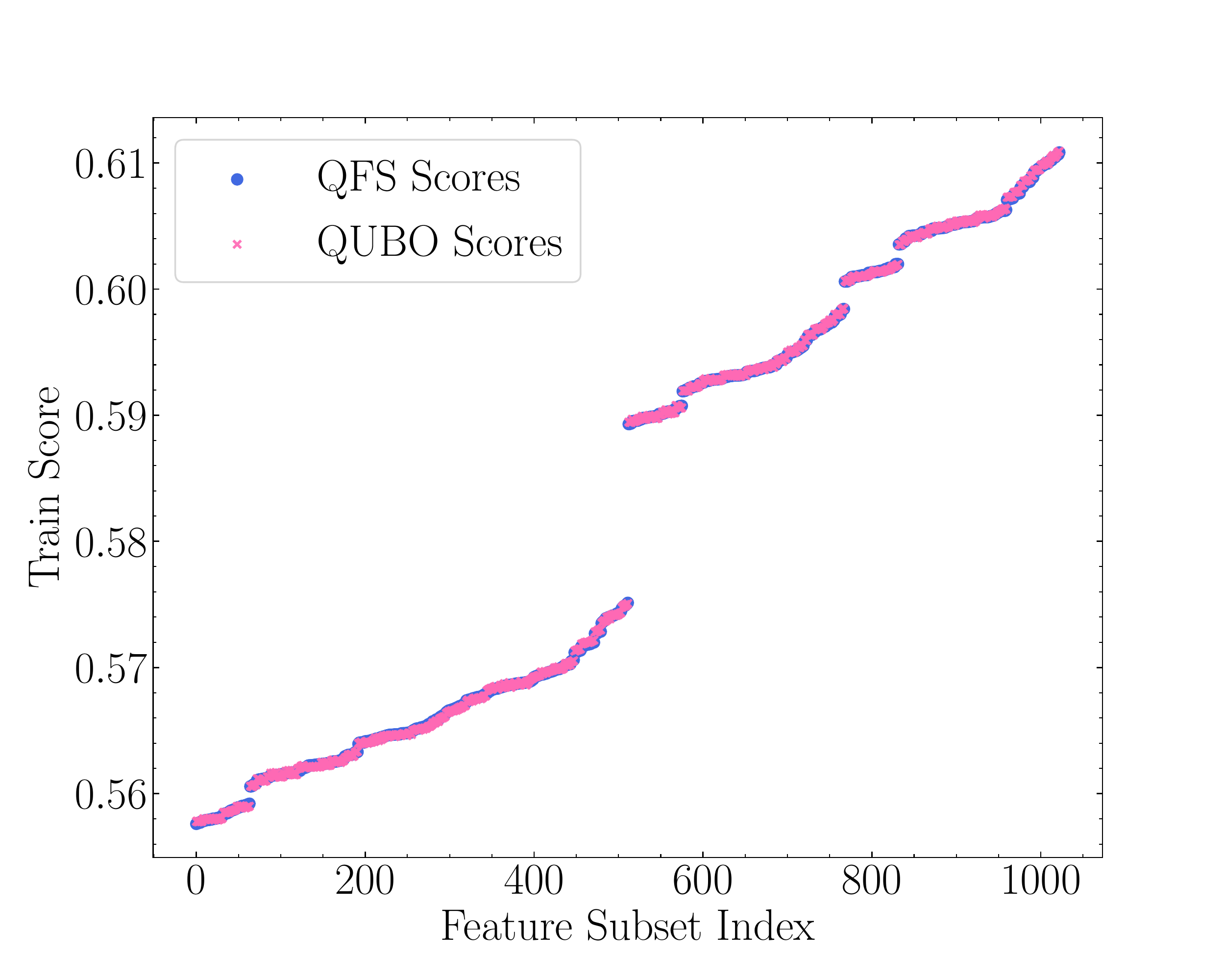}};
\node[anchor=north west] at (1, -5) {\includegraphics[width=0.4\textwidth]{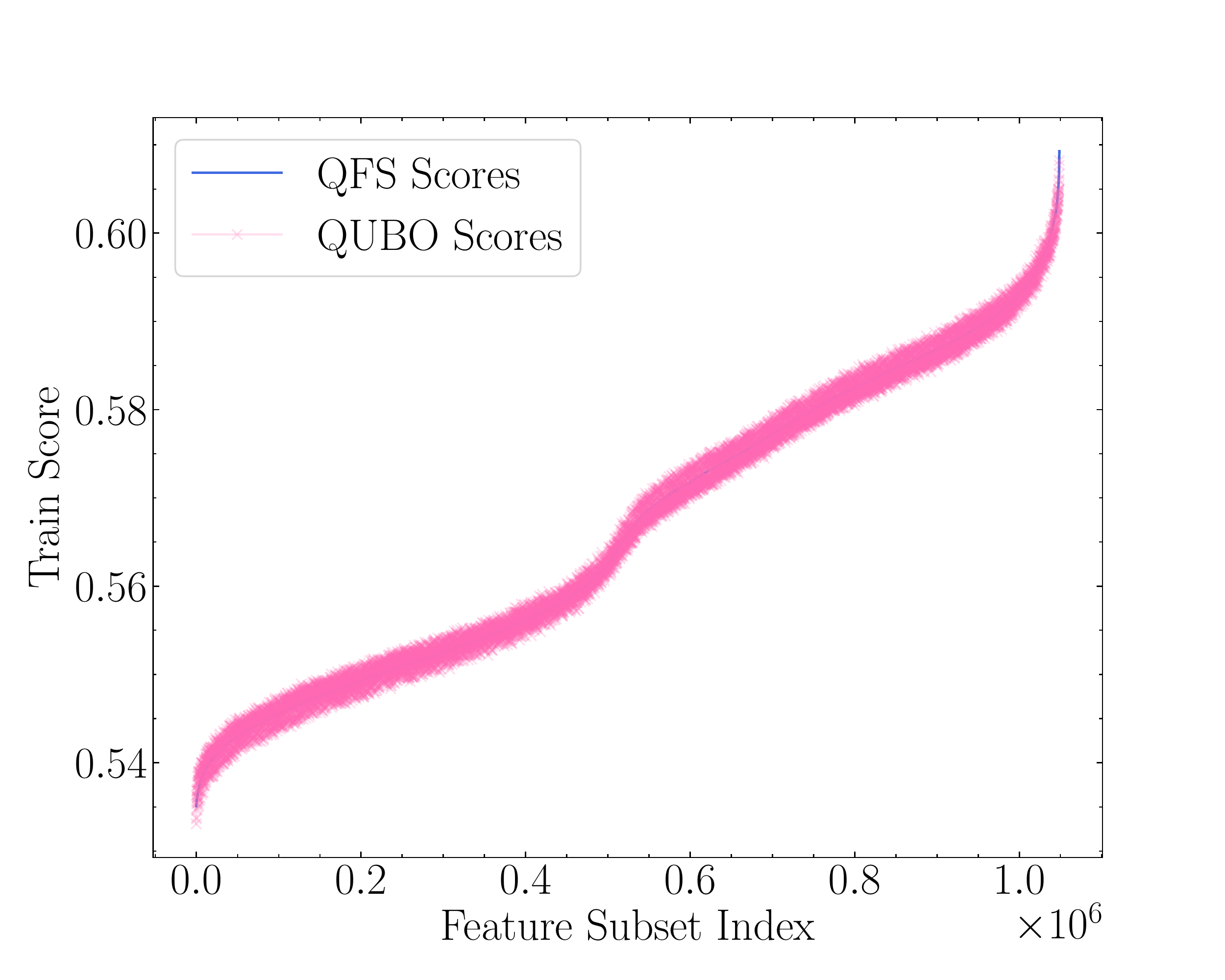}};
\node at (-5, -0.3) {(a)  \small{$5$ features}};
\node at (-5, -5.3) {(c)  \small{$10$ features}};
\node at (2, -0.3) {(b)  \small{$15$ features}};
\node at (2, -5.3) {(d)  \small{$20$ features}};
\end{tikzpicture}
        \captionsetup{singlelinecheck = false, format= hang, justification=centerlast, font=footnotesize, labelsep=space}
        \caption{The figure shows the sorted scores $\mathcal{S}_{\text{train}}\left(\boldsymbol{x}_m\right)$ found using exhaustive search on the VarQFS model as well as the respective scores from the fitted QUBO model. The ${x}$ axis corresponds to the different feature subsets represented by indices.}
     \label{fig:lpfit_bp}
\end{figure}

First, an exhaustive search is employed to find the objective values $\mathcal{S}_{\text{train}}\left(\boldsymbol{x}_m\right)$ for all possible feature combinations.
Then, we define a linear program with $n^2$ variables and $2^n$ constraints and use the $\ell 2$ norm to fit a QUBO with CPLEX \cite{cplex2009v12} for $n=\set{5, 10, 15, 20}$ qubits. 
The mean squared error for the various $n$ is at most of the order $10^{-7}$. 
Thus, we conclude that the individual scores are well approximated for all tested model sizes. This is visualized in Fig.~\ref{fig:lpfit_bp} where we show the scores from the exhaustive search as well as the scores corresponding to the fitted QUBO model.

We would like to point out that these results solely present an empirical indication that the specific feature selection problem investigated in Sec.~\ref{sec:results} may be effectively trained for larger system sizes. However, it remains an open question for future research to generally understand what training data leads to local or global observables.

\bibliographystyle{arxiv_no_month}
\bibliography{bibliography}

\end{document}